\documentclass[10pt, conference, compsocconf]{IEEEtran}

\usepackage[final]{microtype}
\usepackage{graphicx}
\usepackage{subfigure}
\usepackage{booktabs} 
\usepackage{mathtools}
\usepackage{comment}
\usepackage{algorithm}
\usepackage[table,xcdraw]{xcolor}
\usepackage{multirow}
\usepackage{amsmath,amssymb,amsfonts}
\usepackage{algorithmic}
\usepackage{textcomp}
\usepackage{xcolor}
\usepackage{xspace}
\usepackage{fancyhdr}
\usepackage{mathptmx}
\usepackage[normalem]{ulem}
\usepackage{caption}
\usepackage{ragged2e}
\usepackage{makecell}
\usepackage{tikz}
\usepackage[hyphens]{url}
\usepackage[sort,nocompress]{cite}
\usepackage[figurename=Fig.]{caption}
\usepackage[keeplastbox]{flushend}
\usepackage{circledsteps}
\usepackage[bookmarks=true,breaklinks=true,letterpaper=true,colorlinks,linkcolor=black,citecolor=blue,urlcolor=black]{hyperref}

\newcommand{\PCignore}[1]{}

\def\Snospace~{\S{}}

\newcommand{\squishlist}{
 \begin{list}{$\bullet$}
  { \setlength{\itemsep}{0pt}
     \setlength{\parsep}{3pt}
     \setlength{\topsep}{3pt}
     \setlength{\partopsep}{0pt}
     \setlength{\leftmargin}{1.5em}
     \setlength{\labelwidth}{1em}
     \setlength{\labelsep}{0.5em} } }

\newcommand{\squishlisttwo}{
 \begin{list}{$\bullet$}
  { \setlength{\itemsep}{0pt}
     \setlength{\parsep}{0pt}
    \setlength{\topsep}{0pt}
    \setlength{\partopsep}{0pt}
    \setlength{\leftmargin}{2em}
    \setlength{\labelwidth}{1.5em}
    \setlength{\labelsep}{0.5em} } }

\newcommand{\squishend}{
  \end{list}  }

\newcommand{\TODO}[1]{\textcolor{red}{#1 \textit{-TODO}}}
\newcommand{\TK}[1]{\textcolor{blue}{TK: #1}}

\definecolor{amber}{rgb}{0.75,0.35,0.0}
\newcommand{\hl}[1]{\textcolor{black}{#1}}
\newcommand{\rev}[1]{\textcolor{black}{#1}}

\newcommand\circled[1]{\tikz[baseline=(char.base)]{
            \node[shape=circle,draw,inner sep=0.5pt] (char) {{#1}};}}

\newcommand{\alg}{MAGMA\xspace}
\newcommand{\framework}{M3E\xspace}
\newcommand{\group}{group\xspace}

\usepackage{cite}
\usepackage{amsmath,amssymb,amsfonts}
\usepackage{algorithmic}
\usepackage{graphicx}
\usepackage{textcomp}
\usepackage{xcolor}
\usepackage[hyphens]{url}

\title{MAGMA: An Optimization Framework for Mapping Multiple DNNs \\on Multiple Accelerator Cores} 

\author{\IEEEauthorblockN{Sheng-Chun Kao}
\IEEEauthorblockA{
Georgia Institute of Technology \\
felix@gatech.edu}

\and
\IEEEauthorblockN{Tushar Krishna}
\IEEEauthorblockA{
Georgia Institute of Technology \\
tushar@ece.gatech.edu}
}

\begin{document}
\maketitle


\begin{abstract}
As Deep Learning continues to drive a variety of applications in edge and cloud data centers, there is 
a growing trend towards building large 
accelerators with several sub-accelerator cores/chiplets.
This work looks at the problem of supporting 
multi-tenancy on such accelerators.
In particular, we focus on the problem of mapping jobs 
from several DNNs simultaneously on an accelerator. 
Given the extremely large search space, we 
formulate the search as an optimization problem and develop an optimization framework called \framework. In addition, we
develop a specialized optimization algorithm called \alg with custom operators to enable structured sample-efficient exploration. 
We quantitatively compare \alg with several state-of-the-art methods, black-box optimization, and reinforcement learning methods 
across different accelerator settings (large/small accelerators) and different sub-accelerator configurations 
(homogeneous/heterogeneous), and observe 
\alg can consistently find better mappings.
\end{abstract}

\IEEEpeerreviewmaketitle
\section{Introduction}



Accelerators for Deep Neural Network (DNN) models are 
commonplace today in both the cloud and edge. 
As AI workloads continue to drive up the demand for compute, there is a trend towards 
building large accelerators housing several sub-accelerator/arrays (summarized in \autoref{table:highlevel_comparison}). 
Key examples include MCM-based SIMBA~\cite{simba}, wafer-scale Cerebras~\cite{cerebras} or scaled-out  platforms~\cite{cloud_tpu, aimt, tpu_spec}.
Some recent studies have also explored heterogeneous multi-accelerator designs enabled via reconfiguration~\cite{planaria} or 
separate \textit{heterogeneous} sub-accelerators~\cite{herald}.

With the emergence of such platforms, enabling multi-tenancy, i.e., multi-DNN mappings on the an accelerator/ platform, is a natural use-case. Data center workloads often run three categories of inference tasks: vision, language and recommendation, and in each task it involves variants of related DNN models~\cite{park2018deep, anderson2021first}. 
In this work, we target all three use cases and focus on batched-job tasks (jobs launched in bulk without latency constraint but with high-throughput need~\cite{park2018deep, anderson2021first,richins2020missing}), e.g., Google photo auto-editing, image tagging, video processing, and voice processing.

There have been a few recent works looking into the problem of mapping multi-DNN workloads on multiple accelerator cores. 
PREMA~\cite{prema} develops a mapper for multi-tenant language tasks, however targeting single-core accelerator.
AI-MT~\cite{aimt} successfully designs a mapper for homogeneous multi-core accelerators and shows performance improvement over vision and language tasks. Herald~\cite{herald} targets heterogeneous multi-core accelerators and systematically analyzes the benefit of heterogeneity in dataflows across the accelerator cores for AR/VR workloads (vision tasks).
These works demonstrate the impact of a mapping (of DNNs) for the new multi-tenant multi-core accelerators, which is of rising interest, as shown in \autoref{table:highlevel_comparison}. However,
all these approaches rely on manually-designed heuristics. This limits their 
scalability to diverse accelerator back-ends and emerging workloads.
In this work, we develop an automatic mapping search process which includes two specific contributions (i) an optimization framework and (ii) a novel optimization algorithm.


\begin{figure}
\begin{center}
\includegraphics[width=1\linewidth]{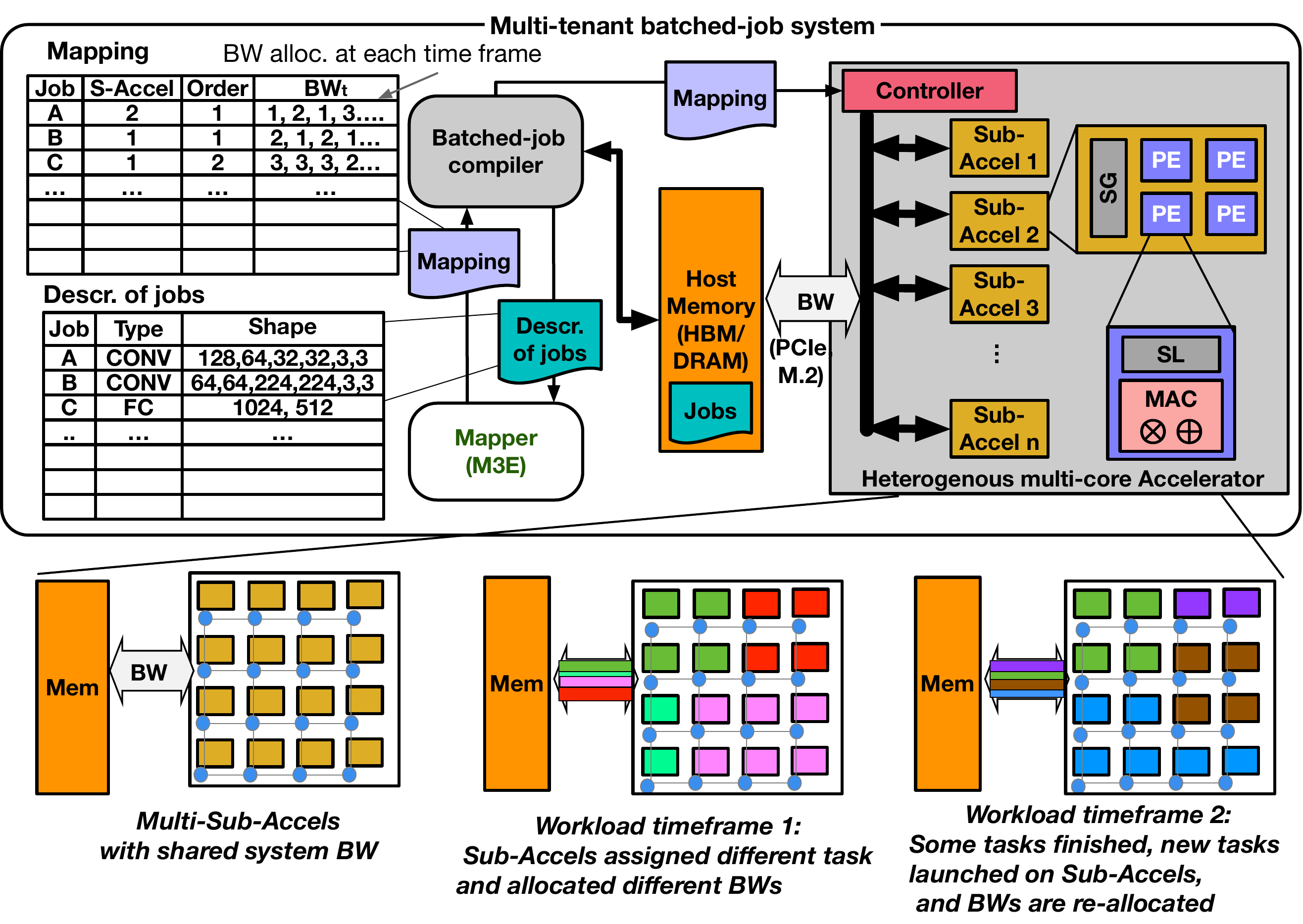}
\end{center}
\vspace{-0.2cm}
\caption{Multi-tenant multi-core accelerator.}

\label{fig:system}
\end{figure}

\begin{figure}
\begin{center}
  \fcolorbox{white}{white}{
\includegraphics[width=1\linewidth]{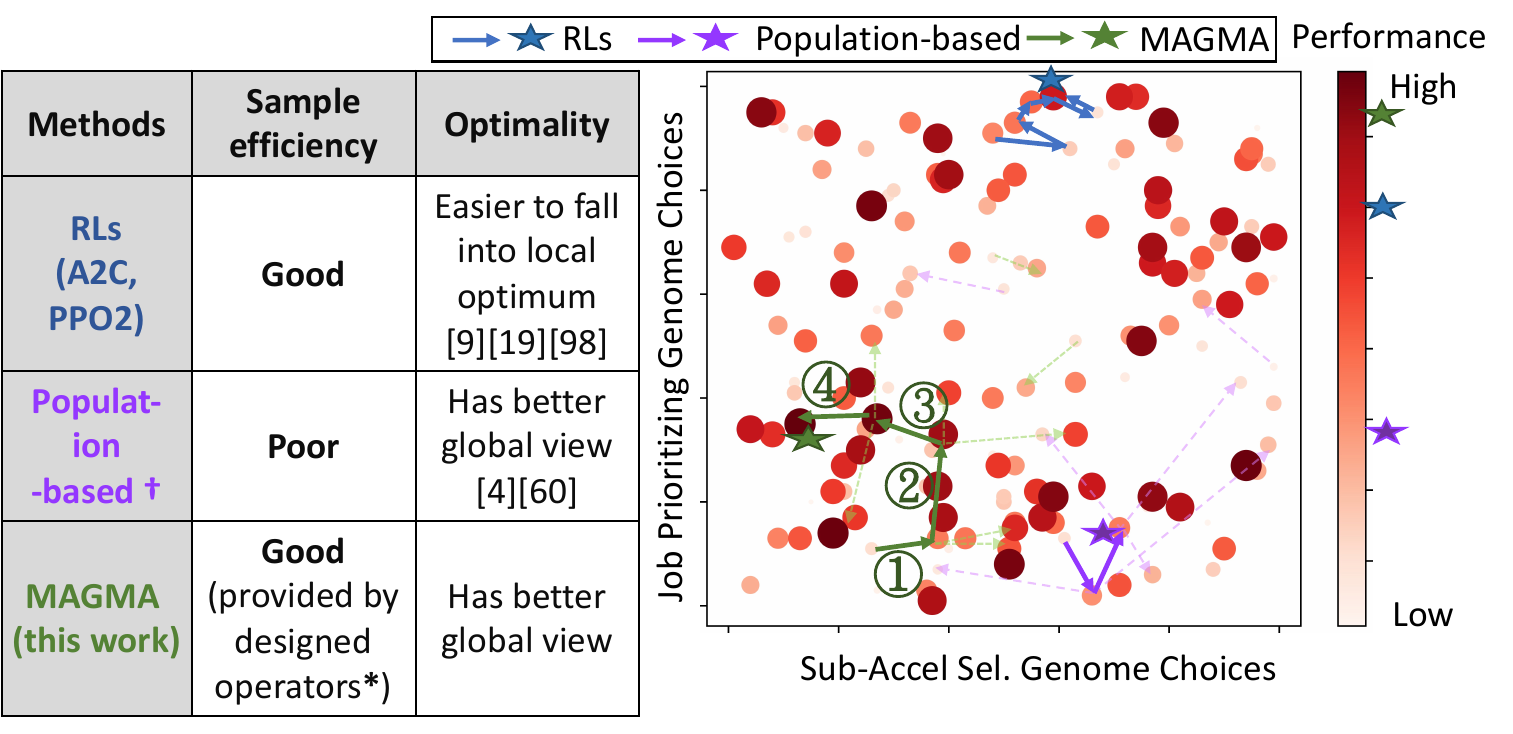}
}
\end{center}
\vspace{-0.3cm}
\caption{\rev{The high-level characteristics of different optimization methods (see \autoref{table:opt_table}) and their effects on their exploration steps~\cite{uberGA, conti2017improving, chang2018genetic,li2011self, bajpai2010genetic}.}} 

\footnotesize{\rev{\textbf{\dag}:Population-based methods: stdGA, DE, CMA-ES, TBPSA, and PSO in \autoref{table:opt_table}}.}

\footnotesize{\rev{\textbf{*}: \textit{Crossover-gen} (MAGMA's operator) can perturb Job-Prior genome (X-axis) while respecting most of the Sub-Accel genome (Y-axis) (step-\circled{2}). Likewise, \textit{Crossover-gen} and \textit{Crossover-accel} (MAGMA's operators) can perturb Sub-Accel genome while respecting most of the Job-Prior genome  (step-\circled{3},\circled{4}). \textit{Crossover-rg} (MAGMA's operator), which respects segments of both genome simultaneously, becomes useful when a segment of Job-Prior genome is highly correlated with corresponding Sub-Accel genes. The designed operators will formulate the exploration in more structured manners, which follow the observed characteristics in this scheduling problem.}}

\label{fig:motivation}
\end{figure}

We propose an optimization framework called Multi-workload Multi-accelerator Mapping Explorer (\framework). In the framework, (i) we develop an efficient encoding scheme to encode the search space of the mapping; (2) we develop several modules to enable the found mapping to effectively orchestrate the data movement across sub-accelerator cores; (3) we enable several commonly used black-box optimization algorithms and two reinforcement learning methods to be leveraged as the underlying optimization methods. In \framework, we break the multi-tenant mapping problem into two components: \textit{sub-accelerator selection} and \textit{job prioritization}. Sub-accelerator selection is where we assign each job a sub-accelerator to execute; job prioritization is where we order the jobs that are assigned to an sub-accelerator.
Each component creates an immense design space by itself. The full design space is the combinatorial probability of both, which becomes as large as O(1e81) (\autoref{sec:search_space}). It also motivates us to design an sample-efficient optimization algorithm.

We propose a custom genetic algorithm-based optimization method for this mapping problem, termed Multi-Accelerator Genetic Mapping Algorithm (\alg)\footnote{This work is available at https://github.com/maestro-project/magma.}. We design custom genetic operators in \alg, which allows it to structurally explore the design space and largely increase its sample efficiency.

Compared to prior work on multi-tenant DNN mapping~\cite{aimt, prema, herald}, this work expands the scope of the problem-space in the following ways:

\squishlist 
\item We utilized optimization-based mapper to solve the mapping problem, while prior arts focus on manually designing a mapper.
\item We target both homogeneous and heterogeneous DNN accelerator platforms.
\item We target a diverse spectrum of 
models across vision, language and recommendation, which exhibit different bandwidth requirements.
\squishend

\noindent Our solution, \framework includes the following novel features:

\squishlist
\item \Circled{1} We frame mapping into an optimization problem and develop key modules to present a complete optimization framework (\framework), which provides the infrastructure for the research community interested in new mapper design or the test-bed for the developed new optimization algorithms. 
\item \Circled{2} A novel optimization algorithm called \alg, which includes several domain-aware operators to enable structured exploration of the large mapping space\rev{, as shown in \autoref{fig:motivation}.} This makes \alg orders of magnitude faster and more sample-efficient than baseline optimization methods~\cite{ga, de, es_origin, cma, pso_paper, tbpsa}, baseline genetic algorithm (GA)~\cite{ga}, and Reinforcement-learning~(RLs)\cite{ppo2, a2c}, as our results demonstrate.
\item \Circled{3} In our evaluation, we found \alg is (geomean) 1.4x and 1.41x better than Herald-like~\cite{herald} and AI-MT-like~\cite{aimt} mappers, and 1.6x better than other comparing optimization methods in homogeneous multi-core accelerators. We found \alg is (geomean) 1.7-2.3x better than Herald-like, 39-52x better than AI-MT-like, 10-13x better than black-box optimizations, and 1.01-1.3x better than RLs in heterogeneous multi-core accelerators.
\squishend

\begin{table}[t]

\centering
\caption{ The comparisons of related works on multi-tenancy and multi-core accelerators. }
\vspace{-0.1cm}
\includegraphics[width=1\linewidth]{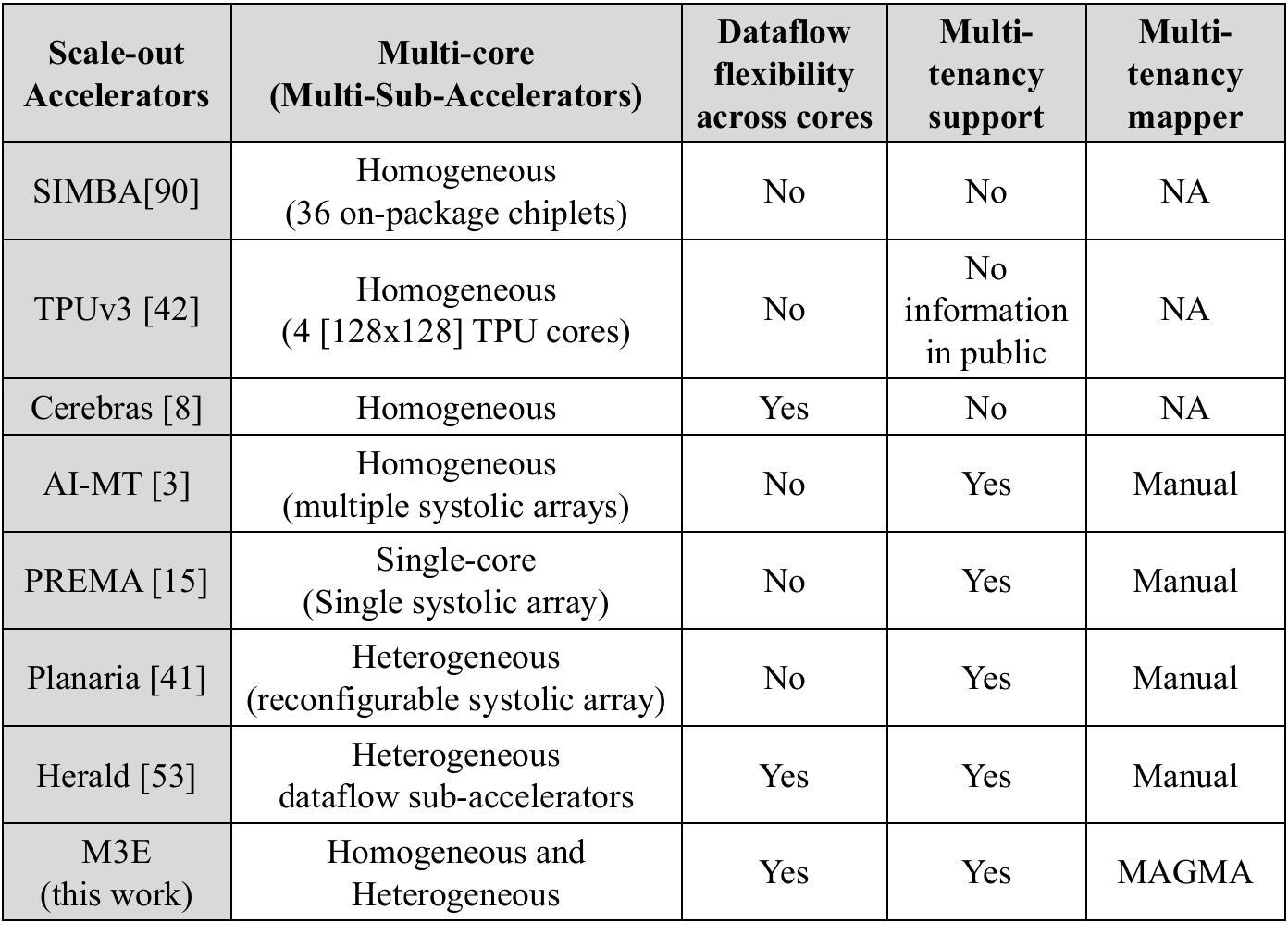}

\label{table:highlevel_comparison}
\end{table}

\section{Background}

\subsection{Characteristics of DNN Models}
In this paper, we consider three types of tasks/ applications that are common in inference data centers~\cite{park2018deep, anderson2021first, richins2020missing} or edge multi-tenant systems~\cite{herald,aimt}: vision, deep recommendation system, and language model.

\textbf{Vision.} Most of vision models~\cite{Resnet,vgg, sandler2018mobilenetv2,szegedy2017inception, Alexnet} are dominated by convolution layers (2D/depth-wise/point-wise) (CONV) and many of them have a MultiLayer Perceptron (MLP) or fully connected layer (FC) at the end~\cite{Alexnet,vgg}.

\textbf{Recommendation.} Recommendation models are either dominated by MLP, attention, or embedding lookup layers~\cite{deeprecsys,dlrm,widedeep, ncf}. With respect to the compute and HW cost, the MLPs and the attention layers are modeled as several FCs. We assume the embedding lookups are kept in the CPU host.

\textbf{Language.} Language models are often dominated by embedding lookup, MLPs, RNNs, and attention layers~\cite{elmo, bert,gpt2, transformerxl}. At long sequence range, attention layers becomes a heavy job owing to its quadratic complexity for both compute and memory~\cite{tay2020efficient,tay2020long}.




\subsection{Multi-core Accelerator}

\subsubsection{\textbf{Accelerator Architecture}}
An accelerator houses multiple cores (termed \textit{sub-accelerators} in this paper).
As shown in \autoref{fig:system}, all sub-accelerators share the ``system BW" via an interconnection network. 
We define system BW as the minimum of main memory (e.g., HBM/DRAM) BW and 
host-to-accelerator 
(e.g., PCIe) BW.
The specific interconnection network architecture can vary depending on the target technology (on-chip~\cite{aimt} versus on-package~\cite{simba} versus wafer-scale~\cite{cerebras}) and the scheduler is agnostic to this.
In this work, we target accelerators with both homogeneous and heterogeneous sub-accelerators. The motivation for heterogeneity among sub-accelerators 
comes from diverse dataflow preferences for different kinds of layers. 
For instance, certain sub-accelerators could be optimized for convolutions~\cite{eyeriss_isca, nvdla} to service vision models, some for GEMM~\cite{tpu, nvidia_volta} to service NLP models, and some for
and embeddings to service recommendation models\cite{hwang2020centaur}. Recent work Herald~\cite{herald} also shows that heterogeneous accelerators are beneficial for multi-DNN workloads. 

\subsubsection{\textbf{Sub-Accelerator Architecture}}
\label{sec:dataflow}

Each sub-accelerator in our system 
is a conventional DNN accelerator that is
comprised of an array of Processing Elements (PE). Each PE has a MAC to compute partial sums, and local scratchpad (called \textit{SL} in this paper) to store weights, activations, and partial sums. Each sub-accelerator also houses a shared global scratchpad (\textit{SG}) to prefetch activations and weights from HBM/DRAM for the next tile of computation that will be mapped over the PEs and SLs. Networks-on-Chip (NoCs) are used to distribute operands from the SG to the SLs and write the outputs back to the SG.

\textbf{Dataflow.}
Given a DNN layer, each sub-accelerator 
employs a dataflow (\textit{aka local-mapping}).
The dataflow determines the loop order, parallelism dimensions, and tile sizes for running the layer on the sub-accelerator.
From the data movement perspective, 
a tile is a basic data movement unit from DRAM/HBM to the SG. 
The tile sizes are bound 
by the size of the SG buffer.
The SG is double-buffered~\cite{maestro, eyeriss_isscc} to 
try and hide the data-fetching latency of the current tile from DRAM/HBM
behind the compute latency. However, \textit{if the bandwidth to main memory is insufficient to hide the fetch latency (based on the bandwidth allocation determined by the scheduler), the accelerator will stall.}
For instance, the NVDLA~\cite{nvdla} dataflow keeps weights in the outermost loop (i.e., weight-stationary) and 
schedules the weight tiles 
spatially over the array along the input and output channels for parallelism.

\section{Problem Formulation}
\label{sec:setup}
The focus of this paper is \textit{designing an automated mapper for multi-tenant DNN accelerators housing several homogeneous or heterogeneous cores}.
The output of our mapper is a \textit{global mapping}, i.e., mapping of independent jobs across the accelerator cores over space and time.
We describe the problem formulation and assumptions in detail here.
In \autoref{sec:framework}, we formulate this mapping problem as an optimization problem and propose an optimization framework. In \autoref{sec:methodology}, we propose an algorithm targeting this mapping problem.

\textbf{Jobs.}
We focus on systems targeting multi-tenancy and batched-job tasks~\cite{park2018deep, anderson2021first, richins2020missing}. Batched-job tasks are usually not latency sensitive and targeting high throughput; in contrast, single-job tasks are often latency-sensitive and targeting minimum response time. Moreover, multi-tenant batched-job tasks also has much less layer dependency issue than single-tenant single-job tasks for following reasons. i) As shown by the observation in AI-MT~\cite{aimt}, multi-tenancy, running multiple neural networks together, can largely alleviate the layer dependency problem as layers from different neural networks can be freely scheduled without any dependency issue. ii) In batched-job tasks, where often hundreds to thousands of activations are running by the same model (e.g., video processing), in practice, the activations will be broken down to mini-batches because of memory concern. These mini-batches are independent with each others, which further alleviate the dependency issue when scheduling these mini-batches. In this paper, we refer to a ``job" as a mini-batch of a layer --- a job is a mini-batch of activations and a set of weight parameters of a layer, where the layer belongs to one of the independent models in the multi-tenant system. 

\textbf{Group.} 
At the host, we run a light-weighted control program to divide a pool of queuing jobs into multiple "dependency-free~\cite{aimt}" \group{s} (Jobs inside each \group have no dependency). The \group size is a hyper-parameter. Larger \group size could increase the search space of the mapper and increase the expected performance; however, larger \group size also increase the difficulty for the mapper which could lead to sub-optimal performance. Also, \group size should be larger or equal to number of sub-accelerators to avoid some sub-accelerators being idle completely.

\textbf{Mapping.} Mapping is the strategy to orchestrate the jobs assignment and job prioritizing across sub-accelerators over space and time. It could also be called \textit{global mapping} (a global control of the data movement across host and sub-accelerators), while dataflow (\autoref{sec:dataflow}) is called \textit{local mapping} (a local control of the data movement inside one sub-accelerator).

\textbf{Mapper.}
The mapper for multi-tenant multi-core accelerators for DNN workload is a comparatively new and rarely explored topic. Recent works such as PREMA~\cite{prema}, AI-MT~\cite{aimt}, and Herald~\cite{herald} have motivated the need of an effective mapper for new DNN accelerators targeting multi-tenant DNNs and multiple sub-accelerator cores. 

A mapping for such accelerators could be a manually-designed heuristic or found by a mapping optimizer. The manual-designed mapping heuristic~\cite{aimt, herald} is often highly tuned for specific accelerators/ systems or tasks/ workloads. The challenge with this approach is that whenever the underlying accelerators or the targeting tasks change, the mapper need to be re-designed or at least undergo another lengthy cycle of engineer-intensive tuning process. 

In this work, unlike the previously mentioned manual-designed mappers~\cite{aimt, herald}, we explore the potential of the mapping optimization method. 
It has the benefit of "automatics", i.e., we simply need to re-run the search of the optimizer whenever the underlying accelerator evolves or target task changes, which happens more and more frequently at the era of the fast-paced hardware evolution. We discuss the proposed mapping optimization framework next.


\vspace{-0.1cm}
\section{Optimization Framework (\framework)} 
\label{sec:framework}
\begin{figure}[t]
\begin{center}

\includegraphics[width=1\linewidth]{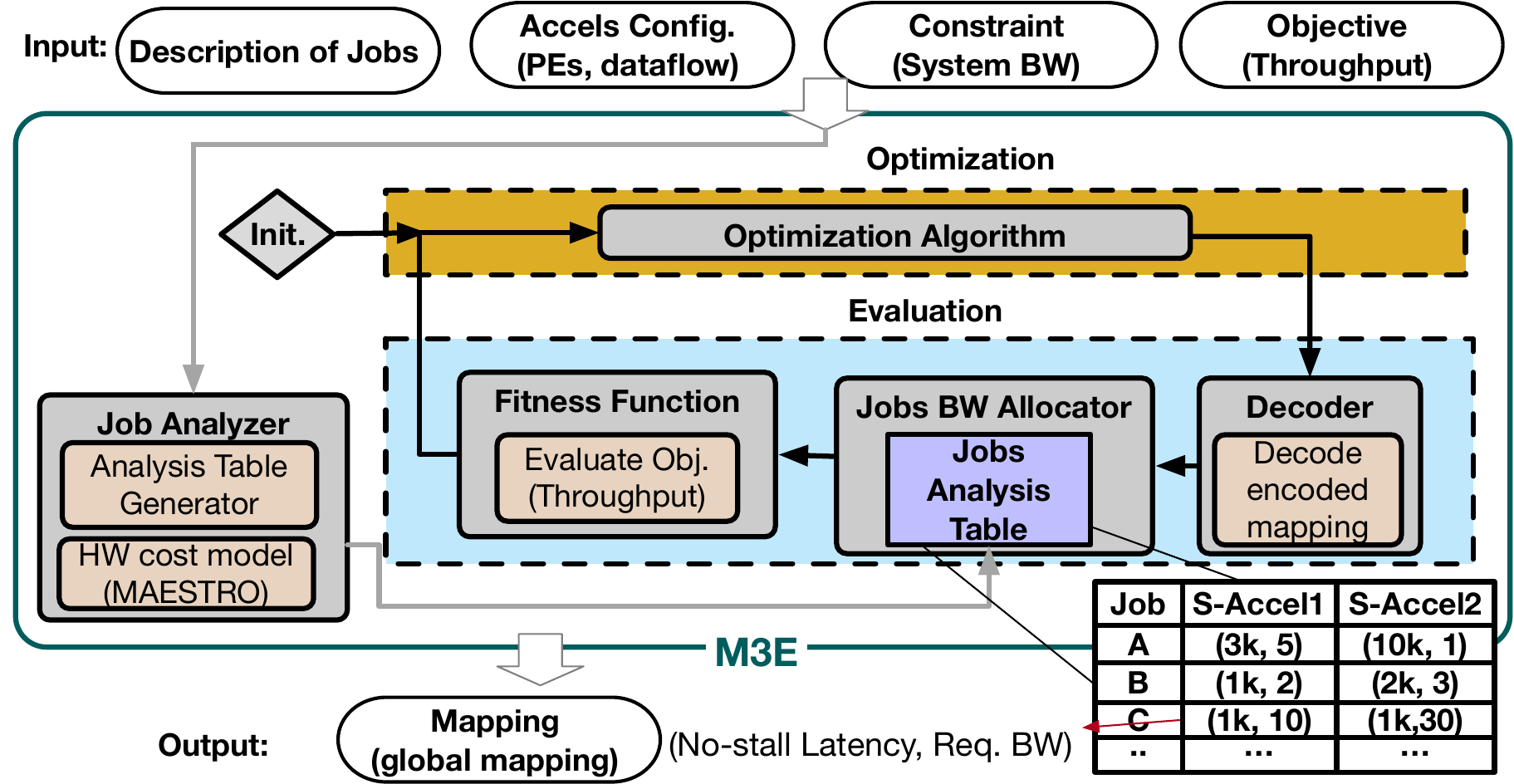}
\end{center}
\vspace{-0.2cm}
\caption{ The structure and flow\textsuperscript{\dag} of \framework. }


\label{fig:framework}
\end{figure}

We propose a mapping optimization framework for multi-tenant heterogeneous accelerator. The structure of our proposed framework called \framework is shown in \autoref{fig:framework}.

At a high-level, the  \framework consists of an optimization phase and an evaluation phase. At evaluation phase, the candidate mapping is evaluated by the cost model. At optimization phase, the optimization algorithm tweak the mapping based on the feedback from the evaluation phase. The optimization-evaluation loop happen iteratively until the targeting objective converges or after a fixed set of time epochs.

The mapping consists of two key components:
\squishlist
\item \textbf{Sub-accelerator selection:} the assignment of each job to a specific sub-accelerator.
\item \textbf{Job prioritization:} execution order of jobs on a given sub-accelerator.
\squishend

To successfully frame a problem into an optimization process, there are two critical pillars: encoding (how the search space is described) and optimization algorithm (how the search space is explored). We describe them as follows.

\subsection{Encoding}
\label{sec:encoding}
An encoded mapping should encode the joint strategy of job prioritization and sub-accelerator selection. We encode them into a series of values with two separate sections, as shown in ~\autoref{fig:ga_combined}(a). The sub-accelerator selection section decides which job goes to which sub-accelerator. The job-prioritizing section decides the order of the jobs in each sub-accelerator. The length of the section is equal to \group size. A full mapping consists of two sections with total length 2x \group size. We describe the encoding using the example in \autoref{fig:ga_combined}(a) assuming two sub-accelerators and a \group size of 5.

\textbf{Sub-accelerator Selection Section.} Each value describes the sub-accel ID for the corresponding job. For example, jobs J1 and J4, are assigned to sub-accel 1, and J2, J3, and J5 are assigned to sub-accel 2 as shown in the sub-accel selection part of the decoded assignment in \autoref{fig:ga_combined}(a).

\textbf{Job Prioritizing Section.} Each value describes the priority of the corresponding job. The priority value ranges from 0 to 1, where 0 is the highest priority. We order the job assigned to a certain sub-accelerator by the order of priority value. For example, J1 runs before J4 in sub-accel 1 as shown in the job prioritizing part of the decoded assignment in \autoref{fig:ga_combined}(a).

\begin{figure}[t]
\begin{center}
\includegraphics[width=1\linewidth]{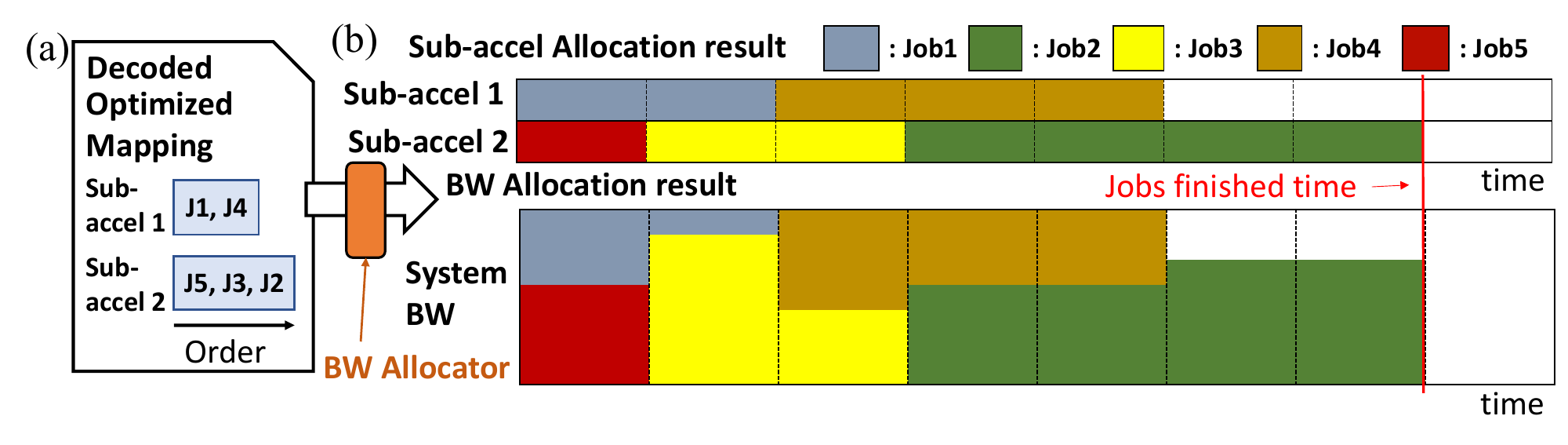}
\end{center}
\vspace{-0.3cm}
\caption{(a) Mapping description from the decoder. (b) The example BW and sub-accelerators allocation results.}

\label{fig:bw_example}
\end{figure}

\subsection{Optimization Algorithms Supported}
With an encoded search space and the support of building blocks of \framework (described later in \autoref{sec:building_block}), we support several popular optimization algorithms, as shown in \autoref{table:opt_table}. We include multiple black-box optimization methods such as Differential Evolution~\cite{de}, Covariance Matrix Adaptation Evolution Strategy~\cite{cma}, and Particle Swarm Optimization~\cite{pso_paper}. In addition, we also include two widely-used reinforcement learning methods --- Advantage Actor-Critic (A2C)~\cite{a2c} and Proximal Policy Optimization (PPO2)~\cite{ppo2}. Finally, we also support our novel optimization algorithm called \alg (\autoref{sec:methodology}). 
With the established framework, \framework can also easily be extended to support other algorithm proposals.

\subsection{Objective and Constraints}

We target throughput as our main objective. However, other objective can also be set (e.g., latency, energy) or formulated (e.g., energy-delay-product, performance-per-watt). The objective can simply be specified as an input to the \framework, as shown in \autoref{fig:framework}. We consider BW constraint (however similarly, other constraints can also be specified). The BW constraints include accelerator-to-host BWs (e.g.,PCIe, M.2) and host memory BW (eg., DRAM/HBM BW), which are common constraints in the practical deployment scenario~\cite{park2018deep, herald}. For simplicity of the optimization framework, we take the minimum of the two (the more stringent BW constraint), as the BW constraint known by the optimization framework, and we name this constraint --- system BW.

\begin{algorithm}[tb]
   \caption{BW Allocator}
   \label{alg:bw_alocator}
\begin{algorithmic}
\footnotesize
   \STATE {\bfseries Input:} Mapping description
    \STATE {\bfseries Output:} $\boldsymbol{BW_{t}^{alloc}}$, t=1,2...T
    \STATE Get $\boldsymbol{Lat_{t}}$, an array of no-stall latency for the parallel jobs at time t, t=0
    \STATE {Get $\boldsymbol{BW_{t}^{req}}$}, an array of required BW for the parallel jobs at time t, t=0
    \STATE {$\boldsymbol{CurJobs_{t}} = \boldsymbol{Lat_{t}} \times \boldsymbol{BW_{t}^{req}}$}
  \WHILE{$\boldsymbol{CurJobs_{t}}$ is not empty}
    \IF{sum($\boldsymbol{BW_{t}^{req}}$) $<$ $BW_{sys}$}
        \STATE $\boldsymbol{BW_{t}^{alloc}}$ = $\boldsymbol{BW_{t}^{req}}$
    \ELSE
         \STATE $\boldsymbol{BW_{t}^{alloc}}$ =$\frac{\boldsymbol{BW_{t}^{req}}\times BW_{sys}}{sum(\boldsymbol{BW_{t}^{req}})}$
    \ENDIF
    \STATE $\boldsymbol{runtimes}$ =$ \frac{\boldsymbol{CurJobs_{t}}}{\boldsymbol{BW_{t}^{alloc}}}$
   \STATE $runtime$ = min($\boldsymbol{runtimes}$)
   \STATE $\boldsymbol{CurJobs_{t}}$ -= $runtime \times \boldsymbol{BW_{t}^{alloc}}$
   
   \STATE $accel_{next}$ = argmin($\boldsymbol{CurJobs_{t}}$)
    \STATE $t$ += $runtime$
   \STATE Fetch the next $Lat$ and $BW^{req}$ of $sub-accel_{next}$, compute $CurJob_{t}$ and insert into $\boldsymbol{BW_{t}^{req}}$, $\boldsymbol{Lat_{t}}$ and $\boldsymbol{CurJobs_{t}}$.
  \ENDWHILE

\end{algorithmic}

\end{algorithm}

\subsection{Building Blocks of \framework.}
\label{sec:building_block}
\subsubsection{\textbf{BW Allocator}}
\label{sec:bw_allocator}
However, in a multi-core accelerator, system BW to the accelerator becomes a global shared resources between cores (sub-accelerators). To evenly allocate the same amount of BW to all the sub-accelerators is an often applied heuristics. However, it will increase the possibility of compute resource under-utilization. E.g., in a normal single-accelerator case, depth-wise CONV jobs are often more memory-intensive than regular 2D CONV jobs, which can make the accelerator under-utilized when running depth-wise CONV while it is fully-utilized when it runs 2D CONV. In the multi-core accelerator, where the system BW is a global shared resources~\cite{park2018deep}, it gives us a chance to reallocate the BW to alleviate the under-utilization problem by proving more BW to core running memory-intensive jobs and proving only adequate BW to cores running compute-intensive jobs, which motivates the BW allocator (Algorithm \autoref{alg:bw_alocator}). The BW allocator is reallocating the BW based on the memory BW intensity of different jobs running on different sub-accelerators.

In detail, receiving the mapping, the BW allocator lookup those jobs' no-stall latency  (\autoref{sec:job_analyzer}) and required BW from the job analysis table (\autoref{sec:job_analyzer}), and allocates the system BW to each sub-accelerator at each time frame with the ratio of their required BWs. It outputs the detailed BW allocation results, as shown in \autoref{fig:bw_example}(b).

\autoref{fig:bw_example}(b), a BW allocation results, as an example, we can tell, jobs J1 and J5 will be launched in Sub-accel-1 and Sub-accel-2, concurrently. Sub-accel-2 will be allocated more BW because it is running a more BW-intensive job. When Sub-accel-2 finishes J5 and launches J3, the BW will be re-allocated to reflect the change of live running jobs in the accelerators, where Sub-accel-1's BW is reduced and reallocated to Sub-accel-2 in \autoref{fig:bw_example}.

\subsubsection{\textbf{Job Analyzer}}
\label{sec:job_analyzer}
The job analyzer takes the jobs description as input and estimates the no-stall latency and its required BW for each sub-accelerator using a cost model (described below) to generate a job analysis table as \autoref{fig:framework} shows.
This table serves as a performance lookup table by the BW allocator (\autoref{sec:bw_allocator}) within the optimization loop.

\subsubsection{\textbf{HW cost model for Sub-Accelerators}}
In \framework, we leverage MAESTRO \cite{maestro_web} as our underlying 
cost model for each sub-accelerator because 
of its ability to support diverse accelerator dataflows and configurations\footnote{In this paper, we explore heterogeneity with the aspect of different specialized DNN accelerators configurations (PEs, buffer size, dataflows). However, \framework is general enough, so that it could also consider generic architectures such as CPUs/GPUs/TPUs by plugging in their cost models.}. It supports most of the common DNN layers such as CONV, depth-wise CONV, and fully connected. Given a DNN layer, a HW resource configuration (PE, SL size, SG size, NoC latency, and BW), 
and a mapping/dataflow strategy, MAESTRO estimates the statistics such as latency, energy, runtime, power, and area. 

\subsubsection{\textbf{Job Analysis Table}}
Job Analyzer profiles each job in the task by the cost model~\cite{maestro_web} and stores the profiling results in the Job Analysis Table. In the optimization process, Job Analysis Table serves as a quick look-up table for fitness evaluation to avoid frequently querying the cost model. The profiling has two main information: no-stall latency and no-stall bandwidth, described next.

\textbf{No-stall Latency.}
We define no-stall latency as the latency 
for running each job on each sub-accelerator, 
assuming it has sufficient memory bandwidth (i.e., not memory-bound).

\textbf{No-stall Bandwidth.}
We define no-stall bandwidth as the minimum bandwidth requirement from each sub-accelerator to make it stay compute-bound, 
not memory-bound.

\subsection{\framework Workflow}
\textit{\textbf{Set-up:}} At the start, the user/host sets up the optimizer by feeding in the jobs descriptions, configurations (number of PEs, dataflow) of each sub-accelerators, the system constraint (system BW), and objective (e.g., throughput). 

\textit{\textbf{Pre-process:}} \textbf{Job analyzer} Job Analyzer prepares the Job Analysis Table, as shown in \autoref{fig:framework}. 

\textit{\textbf{Optimization Loop:}} \textit{Optimization phase:} optimization algorithm updates the encoding mapping based on the feedback from the evaluation block. \textit{Evaluation phase:} \textbf{Decoder} decodes encoded mapping into a mapping description, as shown in \autoref{fig:bw_example}(a). \textbf{BW Allocator} takes in the mapping description and allocates the BW for each sub-accelerator. \textbf{Fitness function} extracts the objective and sets it as fitness value. 

This finishes one loop/ epoch of optimization. The optimization loop stops when \framework reaches the set sampling budget (the  number of allowed sampling data points in a search process).

\subsection{Search Space}
\label{sec:search_space}
The full search space of the proposed framework is the combinatorial combination of the choices for sub-accelerator 
selection and job prioritizing. 
Assuming the accelerator has 4 sub-accelerator and we use the \group size of 60. The size of the design space is $(60!)/(15!)^{4} \times (15!)^{4}=60!=O(1e81)$
which is extremely massive. Therefore the \textit{sample efficiency\footnote{Performance improvement over the number of sampling budget.}} of the optimization methods, which decides the convergent rate, becomes a key factor. We describe our proposed sample-efficient optimization method, next.

\begin{figure*}
\begin{center}
\includegraphics[width=1\linewidth]{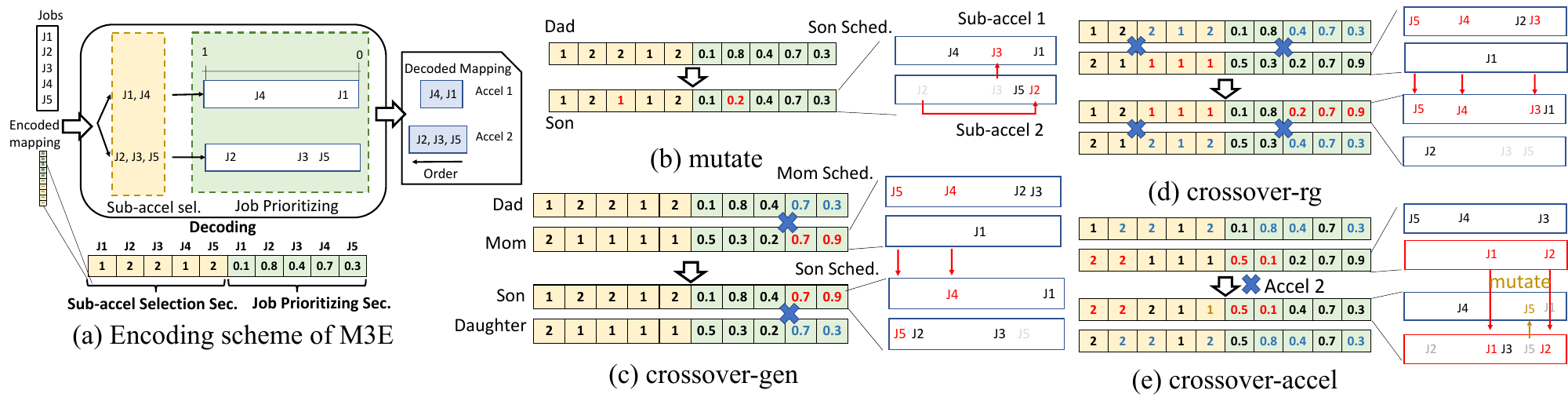}
\end{center}
\vspace{-0.4cm}
\caption{(a) Then encoding scheme of M3E. The genetic operators of MAGMA and their implying update on the mapping: (a) mutation, (b) crossover-gen, (c) crossover-rg, and (d) crossover-accel.}
\vspace{-0.2cm}
\label{fig:ga_combined}
\end{figure*}



\begin{figure}
\begin{center}
\includegraphics[width=1\linewidth]{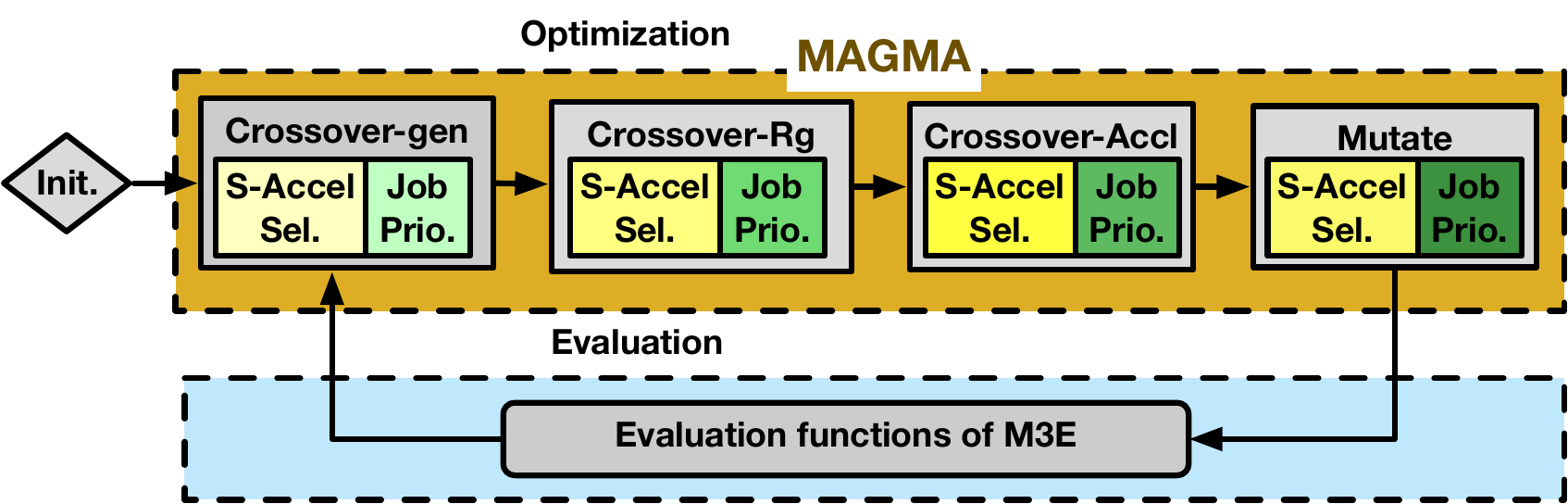}
\end{center}
\vspace{-0.2cm}
\caption{The algorithm flow of \alg.
}

\label{fig:algorithm}
\end{figure}

\section{Optimization Algorithm (\alg)}
\label{sec:methodology}



\alg is a GA-based search technique. 
Its key difference from standard GA is that
it customizes the optimization algorithm's exploration momentum and mechanism (i.e., genetic operators in GA context) for the target search space.


\subsection{Why GA?}

Research shows GA reaches competitive performance with deep reinforcement learning \cite{uberGA,openai_es}, and hyper-parameter optimization problem.  STOKE \cite{schkufza2013stochastic} and Tensor Comprehensions \cite{vasilache2018tensor} use GA to search the space of DNN code optimization. From a search time perspective, GA is light and fast ~\cite{uberGA, openai_es} comparing to many optimizations methods since the optimization mechanism in GA uses simple operations (e.g., crossover and mutations).
A key challenge with standard GA however is that it is not sample-efficient. We address this issue using our customized operators (\autoref{sec:genetic_operator}).

\begin{table}[t]

\centering
\caption{Terminology used in \alg Algorithm.}
\vspace{-0.1cm}
\includegraphics[width=1\linewidth]{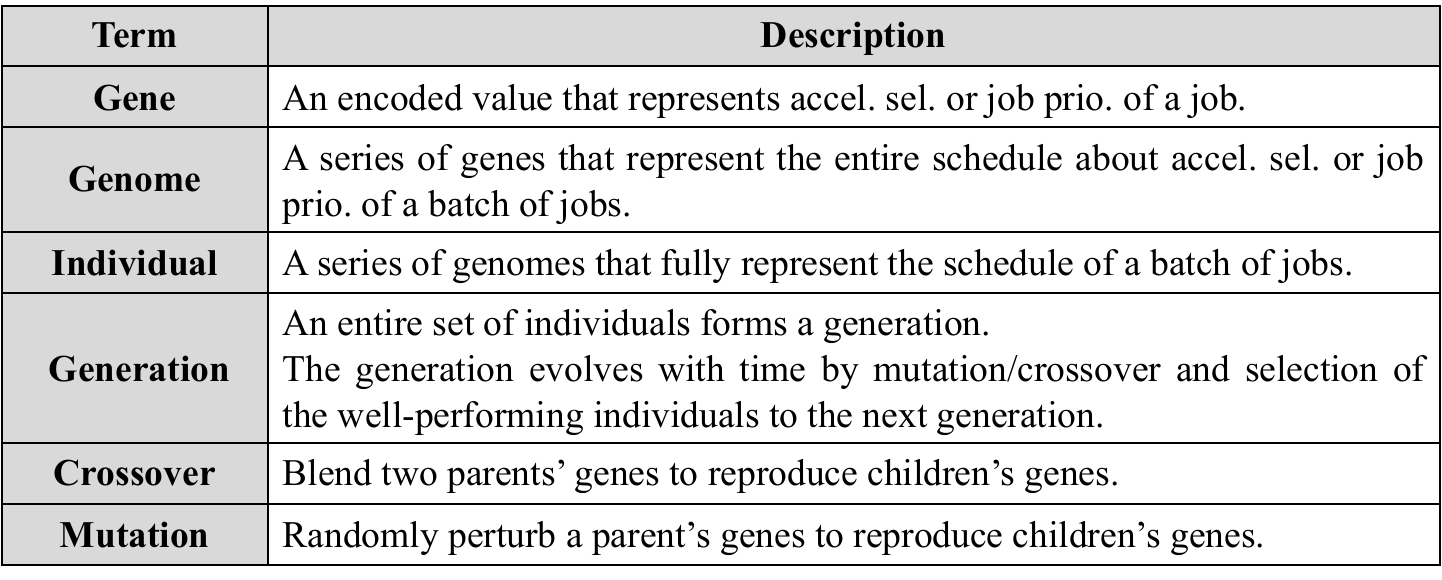}

\label{table:ga_terminology}
\end{table}

\subsection{\alg Algorithm Details}

\subsubsection{\textbf{Terminology and Basics of GA}}
We list the common terminology of GA we use throughout the paper in \autoref{table:ga_terminology}, namely \textit{gene, genome, individual, generation}.
The basic mechanism in GAs is to create a 
population of individuals in each generation.
All individuals are evaluated and sorted based on their fitness. The best performing individuals are used as parents to create a new population of individuals using 
genetic operators (~\autoref{sec:genetic_operator}). The goal of GA is to perturb genes (i.e., components of the schedule) and retain well-performing ones across generations.

\subsubsection{\textbf{Genetic operators}}
\label{sec:genetic_operator}

\textbf{Standard GA Operators.} The standard genetic operators in GA consist of mutation and crossover. The standard mutation operator randomly mutates some genes. The standard crossover operator samples a pivot point and exchanges the genes of parents according to that pivot point. The sampling efficiency of the GA relies on the efficiency of the genetic operators to sample high-quality next generation.

\textbf{\alg Operators.} In \alg, we inherit the standard mutation mechanism and design \textit{three specialized crossover genetic operators}. \hl{Different crossover operators are designed to preserve different dependency of genes while exploration.} They allow us to explore the scheduling problem in a more strategical manner. We describe the genetic operators next.



\textbf{Mutation.}
During mutation, we randomly select multiple genes (according to the mutation rate) and mutate them to random values. 
\autoref{fig:ga_combined}(b) shows an example when mutating at the third and second genes of two genomes respectively. On the right side of the figure, it shows how the son's genes/schedule are generated by the dad's mutation. J3 is moved to sub-accel 1 because of the first mutation. J2 is moved to a higher priority in sub-accel 2 because of the second mutation. In our experiments, we use a mutation rate of 0.05.

\textbf{Crossover-gen.}
This is a genome-wise crossover. First, we randomly sample a type of genome to crossover. Next, we randomly sample a pivot point and exchange the genes of the genomes. 
There are two benefits of genome-wise crossover. First, we keep the perturbation to the level of the genome, which potentially keeps the good characteristics of the other un-touched genomes, and therefore is more stable throughout the evolution. Second, we eliminate the order dependency of the genomes. The genomes are independently representing their features, where the order of them provides no information (\hl{, i.e., representing Sub-accel Sel. genome first and Job Prio. Genome later does not make the J5 of Sub-accel Sel. and J1 of Job Prio. strongly correlated despite their being next to each other.}). Therefore, a genome-wise crossover, \hl{which operates genomes independently}, enables us to perturb the gene without unnecessary assumptions of the genome order. 
Crossover-gen becomes the major crossover function, which we set the crossover rate as 0.9. 

\autoref{fig:ga_combined}(c) shows an example that we pick the second genome \hl{(Job Prio.)} as the crossover region and the third location of the region as the pivot point. \hl{With the respect of schedule change after crossovering}, in the example, the orders of J4 and J5 in mom's schedule are passed to son's schedule.

\textbf{Crossover-rg.}
This is a range crossover mechanism structured to preserve the \hl{the dependency of genes across genomes}. 
For example, 
\hl{in \autoref{fig:ga_combined}(a), the first and the sixth genes are dependent, since they are both representing some features for J1}.
\hl{We randomly pick a range of genome (e.g., the 3rd to the 5th locations of each genome) and simultaneously crossover all the genes falling into the picked region from both genomes, and thus the cross-genome dependency is preserved. }
With the respect of scheduling change after crossovering, the order and accel selection of J3, J4, and J5 are exchanged between two individuals, as shown in \autoref{fig:ga_combined}(d). Crossover-rg has crossover rate of 0.05. 

\textbf{Crossover-accel.}
This is a crossover method to
preserve the dependency of job ordering within an sub-accelerator.
We randomly select a sub-accelerator and pass the job ordering information of this sub-accelerator to the children. For example,
in \autoref{fig:ga_combined}(e), we select sub-accel 2. Next, \hl{we check the Sub-accel Sel. genome of Mom, copy the genes related to sub-accel 2 (the first and second genes of both genomes in (e)), and paste them to son's genomes.}

To increase load balancing, the original jobs assigned to sub-accel 2 in Son will be randomly mutated. Crossover-accel has crossover rate of 0.05. 

%

\subsubsection{\textbf{Hyper-parameter Tuning}}
The above mentioned mutation, crossover rates, populations sizes, and elite ratios are hyper-parameters in \alg. We applied a hyper-parameter search via a Bayesian optimization framework~\cite{bergstra2013making} to select a set of hyper-parameters that makes \alg achieve the highest performance across multiple workloads.


\subsection{Warm-Start of \alg}
\label{sec:transfer}


In this section, we present the techniques we implement to enable the warm-start of the algorithm. Warm-start is a well-known technique in black-box optimizations to enable faster convergence or reaching better objective value. Warm-start works as follows. There are series of tasks to be solved by the optimization algorithms. If the current task is the same or similar to the previous solved tasks, we can take the previous solution to initialize the algorithms. We also implement an warm-start engine, which recognize if the current task fall within the same types of tasks (Vision, Recommendation, or Language), i.e., whether it is similar to the previous solved task. If tasks are within the same type of task, the warm-start engine will take over the initialization job from Init engine (initializing algorithm randomly). In our experiment (\autoref{sec:transfer_exp}), we found warm-start is a useful add-on technique in \alg.

\begin{table}[t]

\centering
\caption{Accelerators configurations/ settings of the experiments.}
\vspace{-0.1cm}
\includegraphics[width=1\linewidth]{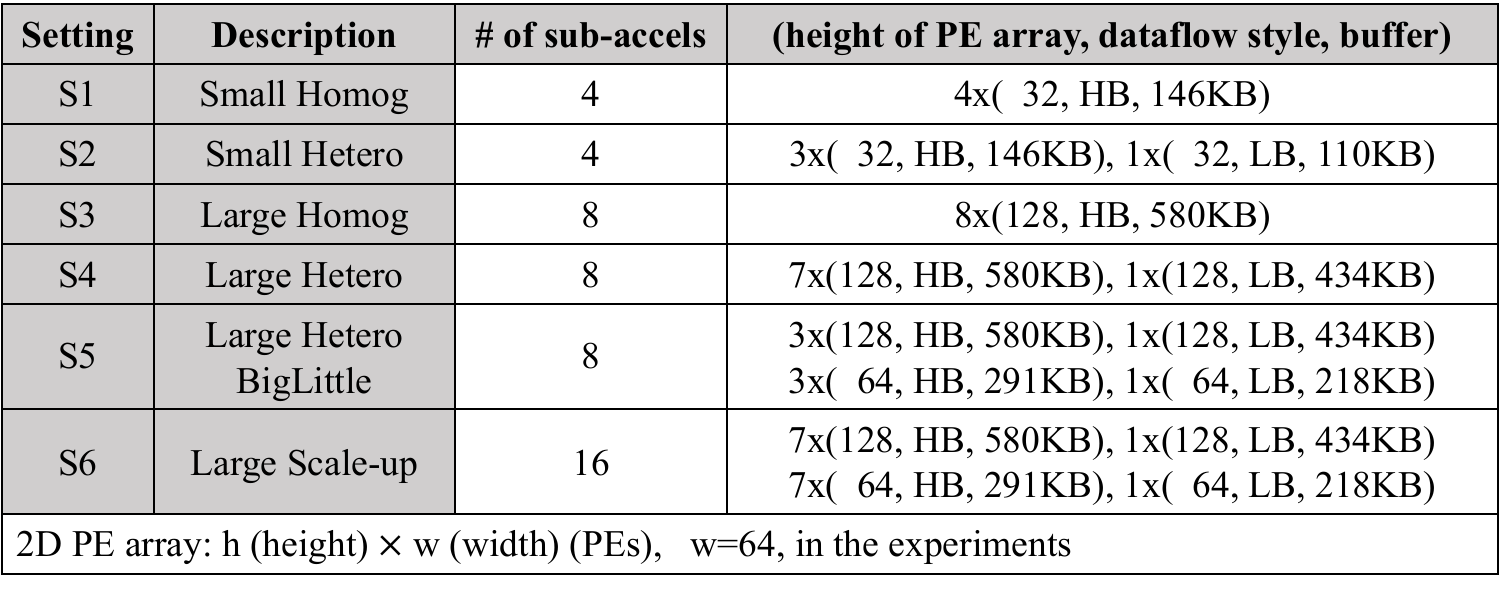}

\label{table:setting}
\end{table}

\begin{table}[t]

\centering
\caption{Supported optimization algorithms in \framework.}

\includegraphics[width=1\linewidth]{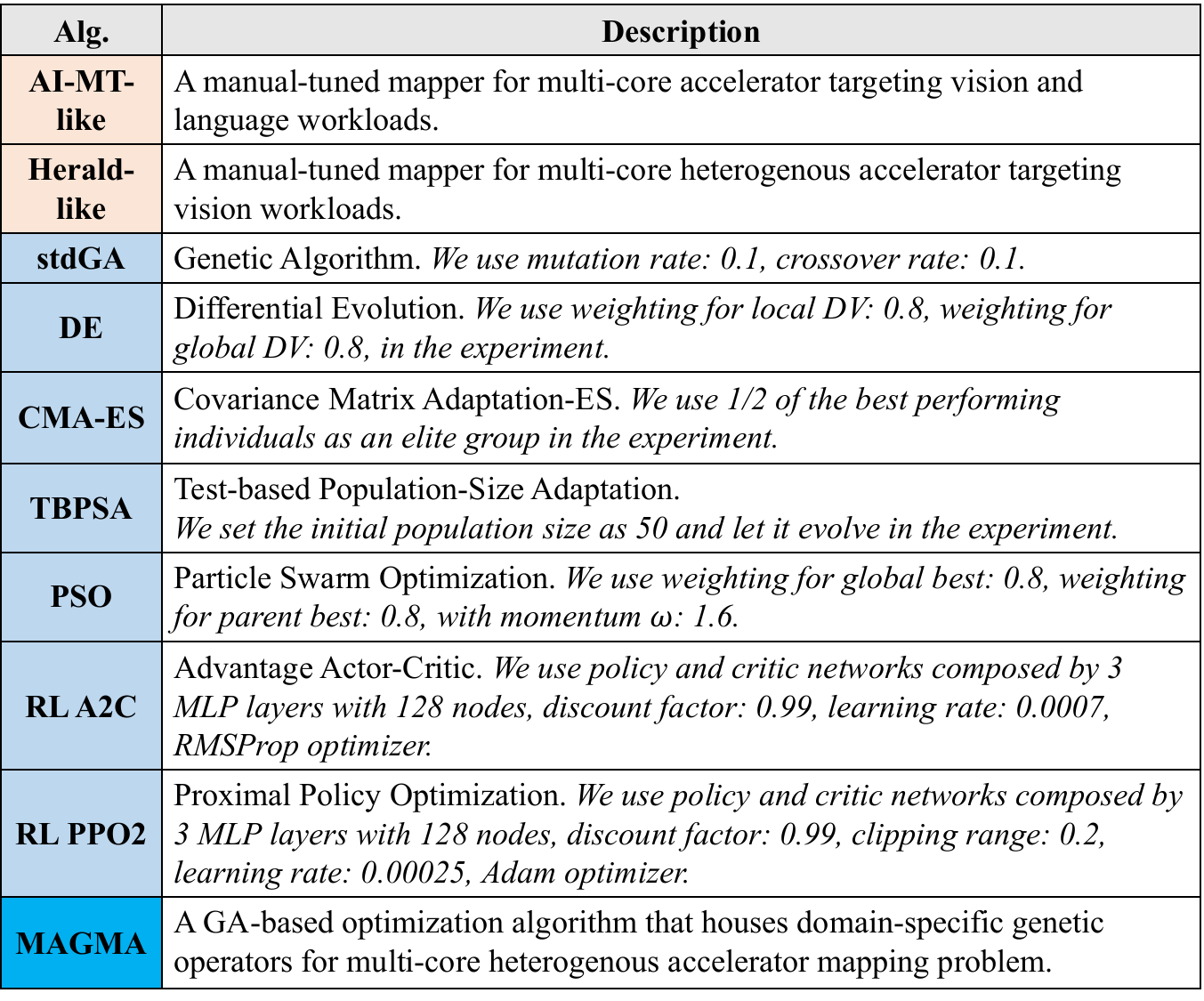}

\label{table:opt_table}
\end{table}

\section{Evaluations}
\label{sec:evaluation}

\subsection{Methodology}

\subsubsection{Target DNN Models}
We consider three different types of tasks/ applications with their corresponding models collected from PyTorch: Vision (\cite{Googlenet,mnasnet,sandler2018mobilenetv2,Resnet,zhang2018shufflenet,squeezenet,vgg}), Language (\cite{bart,bert,camembert,ctrl_lang,electra,flaubert,gpt2,reformer,roberta,t5,transformerxl,xlm,xlmroberta,xlnet}), and recommendations (\cite{din,dien,dlrm,widedeep,deeprecsys}).
The models are the ones used for the batched high-throughput applications (e.g., photo auto-editing, image tagging, and video / voice processing) that we are targeting.

\subsubsection{Task and Benchmark}
We categorize the jobs into 
Vision, Language (Lang), Recommendation (Recom), and Mix (a complex tasks with vision, language, and recommendation model involved simultaneously), four different types of tasks.
We build a benchmark including different tasks motivated by the Facebook's inference accelerator jobs~\cite{park2018deep, anderson2021first}, edge data centers for AI applications~\cite{richins2020missing}, Herald~\cite{herald}. AI-MT~\cite{aimt}, and others~\cite{shen2019nexus}. In the benchmark, we collect models from the four different types of tasks and create several workloads. Each workload contains hundreds to thousands of jobs (one job include a batch of activations and weight parameters of a layer). We chopped them into several ``dependency-free" \group similar to prior works~\cite{aimt}. The objective of the optimization algorithm is to execute these \group with the highest possible throughput. We set the default \group size to be 100 but also study the effect of \group size in \autoref{sec:eval_deepdive}.



\subsubsection{Accelerators}
We consider two classes of accelerator: Small and Large.
For each class, we consider multi-core homogeneous and heterogeneous accelerator settings with different PEs, dataflow, and on-chip buffer.
We construct six different multi-core accelerators, motivated by \cite{herald, aimt, tpu, simba}, as our test-bed in \autoref{table:setting}. S1 and S3 represent homogeneous accelerators. S2, S4-6 represent heterogeneous accelerators. The accelerators are modeled with MAESTRO~\cite{maestro_web}. We uniformly set one dimension of the 2D PEs array to 64\footnote{Based on our observation, most of the popular models that we collected, especially language and recommendation ones, are manually designed to have the tensor shape formed by the multiples of 64. Setting one dimension to 64, which aligns with the tensor shape, ensures higher utilization rate.} and scale the PEs array size by increasing the other dimension.
We consider three kinds of PEs configuration: 32 $\times$ 64 for Small accelerator~\cite{zhu2019energy,fu2020soft,mseddi2019intelligent,du2020new,li2020intelligent}, 64 $\times$ 64 and 128 $\times$ 64 for Large accelerator.
The dataflow style (discussed next) and target tile sizes
determine the buffer sizes for both SL and SG~\cite{maestro}.

\begin{figure}
\begin{center}
\includegraphics[width=1\linewidth]{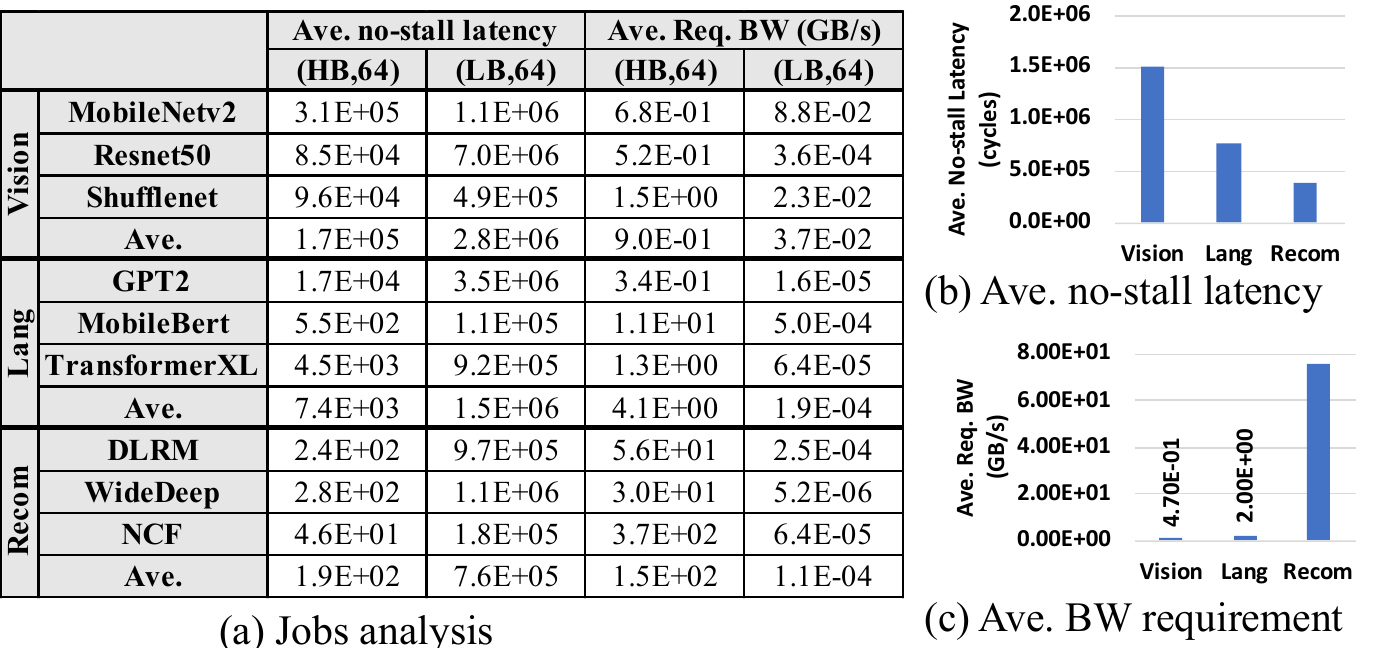}
\end{center}
\vspace{-0.2cm}
\caption{ (a) The average per-job no-stall latency and required BW for no-stall across different models on high (HW) and low (LB) bandwidth mapping style. (b) Average no-stall latency and (c) average BW required for no-stalls across all involved jobs. 
}

\label{fig:analysis_combine}
\end{figure}

\textbf{Sub-Accelerator Dataflow Styles.} For our evaluations, we pick 
two distinct dataflow styles for the heterogeneous sub-accelerators: High Bandwidth usage dataflow style (HB) (inspired by NVDLA)~\cite{nvdla}) 
and relatively Low Bandwidth usage dataflow style (LB) (inspired by Eyeriss~\cite{eyeriss_isca}).
The HB-style parallelizes across channel dimensions, and shows high-efficiency on late layers for CNN-based (vision) models, while the LB-style parallelize across activations dimensions and excels on the early layers of CNN-based models~\cite{maestro}. For Language and Recommendation, 
we found the HW-style is more compute efficient but
BW intensive, while LB-style is less compute efficient but 
also less BW demanding (\autoref{fig:analysis_combine}). 
Therefore we house both these sub-accelerators 
in a BW constrained accelerator 
platform to act as a good test for our 
optimizer to learn and exploit 
their difference. \framework is general enough 
to run with any heterogeneous combination of two 
or more accelerator styles.

\textbf{System BW.}
The accelerators are executing under frequency 200MHz and bit-width 
of 1 Byte. For the system BW, at the Small accelerator, we consider the BW to be range from 1GB/s to 16GB/s, which is the range of DDR1-DDR4 BW~\cite{ddr_bw} and PCIe1.0 - PCIe3.0~\cite{pcie_spec} BW; at the Large accelerator, we consider the BW to be range from 1GB/s to 256GB/s, which is the range of DDR4-DDR5 ~\cite{ddr5_spec} and HBM BW ~\cite{hbm_spec} and PCIe3.0 - PCIe5.0 and upcoming PCIe6.0 BW~\cite{pcie_spec}. 

\textbf{Evaluation Metric}
In all experiments, we use throughput as \rev{the optimization} objective.

\subsection{Mapper Settings.}

\textbf{Baseline Manual-tuned Mapper.} We use the mapper from Herald~\cite{herald} and AI-MT~\cite{aimt} as the baseline methods (Herald-like and AI-MT-like). Note that the mapper in Herald~\cite{herald} is manual-designed targeting multi-core heterogeneous system with Vision tasks, and the mapper in AI-MT is manual-designed targeting multi-core homogeneous system with Vision and Language tasks. In our evaluation, we also tested their performance in Recommendation and Mix tasks and on both homogeneous and heterogeneous accelerators.

\textbf{Optimization methods.} We enable many commonly-used optimization methods in \framework.
The specific hyper-parameters settings are listed in \autoref{table:opt_table}. For fair comparisons, all optimization methods are given the same sampling budget, 10K data points. Note that batched-job tasks are not latency sensitive, where the jobs and optimization process are executed off-line. Therefore the search time of different methods is not our main concern; instead, the objective of these tasks are to utilize the underlying hardware as efficient as possible, i.e, maximizing the throughput of the underlying hardware.

\textbf{\alg.}
\alg is also one of the optimization methods. We set the number of individuals in a generation, to be as large as \group size. We also constraint \alg to have the same 10K sampling budget, and use population size of 100, and thus have 100 epochs for optimizing.
As for search time, we run the experiments on a desktop with Intel i9-9820 CPU. \alg takes about 0.25 seconds per epoch, and 25 seconds for a full optimization process.

\begin{figure*}
\begin{center}
\includegraphics[width=1\linewidth]{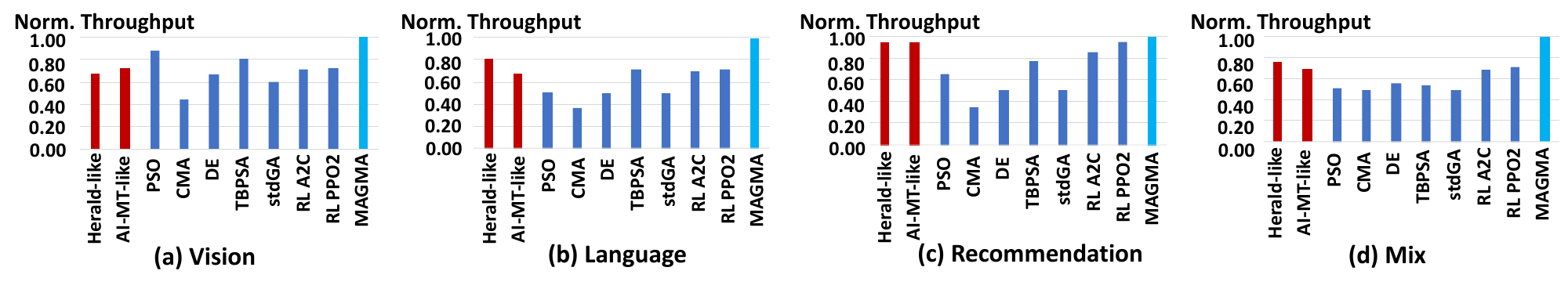}
\end{center}
\vspace{-0.3cm}
\caption{The experiment results on multi-core homogeneous small accelerator (S1) with BW=16 across four tasks. Throughput values are normalized by the value of \alg. \rev{The absolute throughput values of MAGMA in (a-d) are: 249, 397, 194, and 329 GFLOPs.}}
\vspace{-0.2cm}
\label{fig:exp_s1}
\end{figure*}

\begin{figure*}
\begin{center}
\includegraphics[width=1\linewidth]{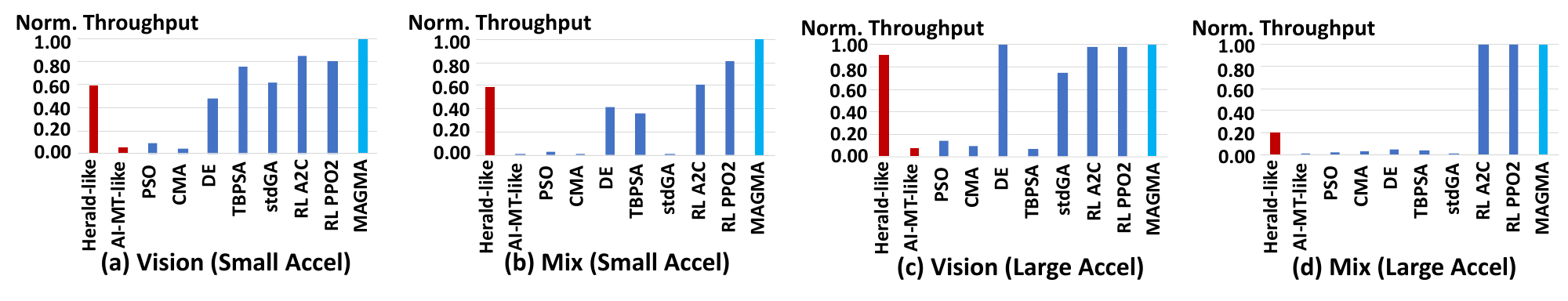}
\end{center}
\vspace{-0.3cm}
\caption{The experiment results on multi-core heterogeneous (a)(b) small (S2, BW=16) and (c)(d) large (S4, BW=256) accelerator on Vision and Mix tasks. Throughput values are normalized by the value of \alg. \rev{The absolute throughput values of MAGMA in (a-d) are: 254, 271, 254, and 383 GFLOPs.}}
\vspace{-0.2cm}
\label{fig:exp_s2_4}
\end{figure*}

\begin{figure}
\begin{center}

\includegraphics[width=1\linewidth]{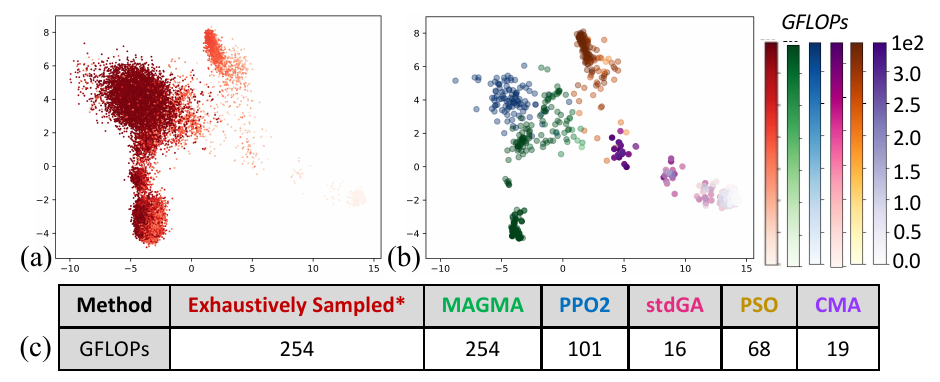}

\end{center}
\vspace{-0.2cm}
\caption{\rev{(a) The full map space, and (b) the 
explored map space and (c) the reached performance of different methods in problem (Mix, S2, BW=16). The axes of (a-b) are 2-dimensional projection by PCA~\cite{pca}. From (b)(c), we can observe CMA, PSO, stdGA, and PPO2 converge to different local optima. MAGMA globally samples a wide region at the start (characteristics as a GA-based algorithm~\cite{bajpai2010genetic}) and quickly converges to an optimum with better performance than other methods (sample efficiency provided by the the designed operators).}}

\footnotesize{\rev{\textbf{Exhaustively sampled*}: Running random sampling for around 2 days and more than 1 million samples collected. It represents the best-effort optimum point. (Note that all other methods are run with the set 10K sampling budget.)}}

\label{fig:scatter_plot}
\end{figure}

\begin{figure}[t]
\begin{center}
\includegraphics[width=1\linewidth]{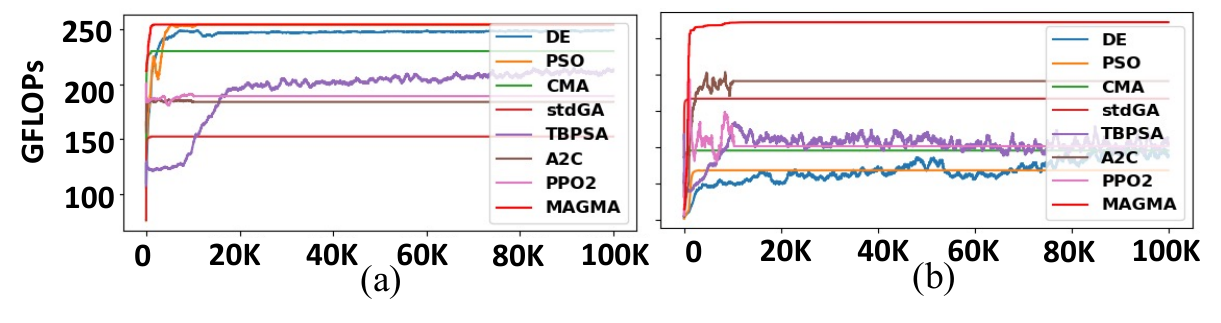}
\end{center}
\vspace{-0.4cm}
\caption{\rev{The convergence curve of different methods across 100K samples on (a) (Vision, S2, BW=16) and (b) (Mix, S3, BW=16). Most of the methods converge before the set 10K sampling budget in (a)(b) (and all experiments in this paper). For few cases, some algorithms require more sampling budget to converge, and we show one of them in (a): TBPSA requires around 20K samples. However, in the end, they all converge to some lower performance points than the points found by MAGMA.}}

\label{fig:convergent_curve}
\end{figure}

\subsection{Latency-BW Characteristics of DNNs}
\label{sec:traffic_analysis}
We start by showing the latency characteristics and bandwidth requirements of the DNN models 
from the three types of tasks when running by itself on two separate dataflow styles (HB and LB). 
We show three of the models from each type of tasks and the average across all the models in that type of tasks in \autoref{fig:analysis_combine}(a). The average values across all model across both accelerators are plotted in \autoref{fig:analysis_combine}(b-c).
From \autoref{fig:analysis_combine}, in general, we can see that the per-job latency of the Vision models is higher because more compute is needed in the CONV dominant models. However, CONV is generally less memory-bound than FC. The data also shows that usually Vision has the lowest BW requirement, and Recommendation has the largest. 


\begin{figure}[t]
\begin{center}
\includegraphics[width=1\linewidth]{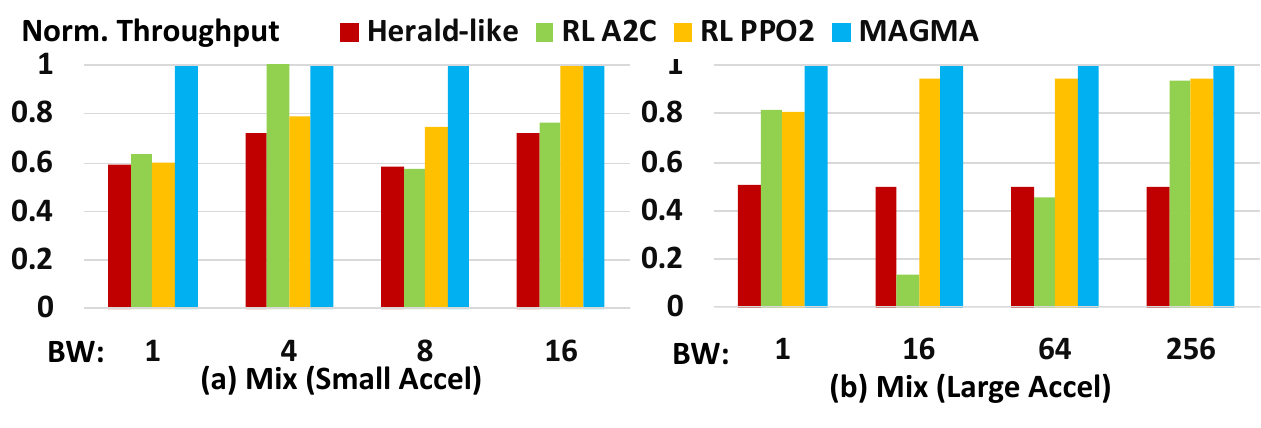}
\end{center}
\vspace{-0.4cm}
\caption{Performance comparisons on multi-core heterogeneous (a) small (S2) and (b) large (S4) accelerator on Mix tasks, given different BWs. Throughput values are normalized by the value of \alg.}

\label{fig:bw_sweep}
\end{figure}

\begin{figure*}[t]
\begin{center}
\includegraphics[width=1\linewidth]{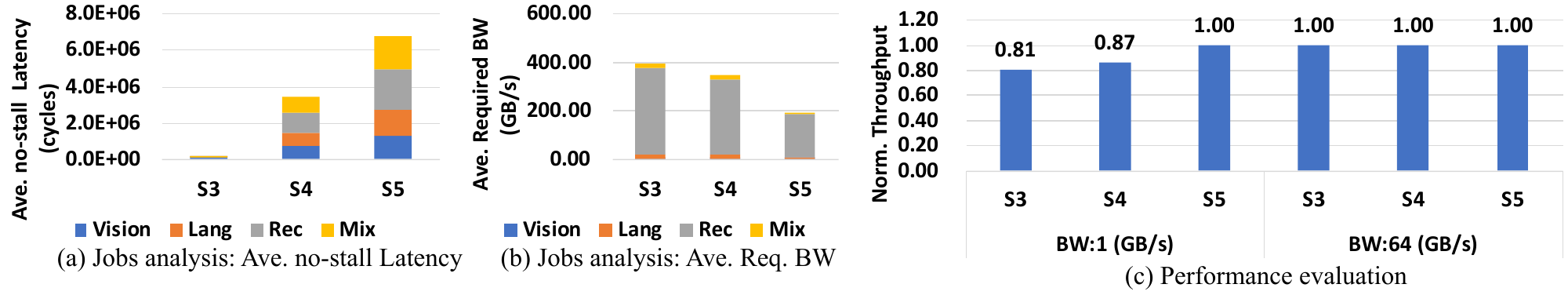}
\end{center}
\vspace{-0.4cm}
\caption{Jobs analysis of the averaged per-job (a) no-stall latency and (b) required BW. (c) Performance evaluation of \alg on S3, S4, and S5 with different BW. In (a-b), we concatenate the four independent no-stall latency (averaged required BWs) into a stacked bar and show the total values. In (c), throughput values are normalized by the value of S5.}
\vspace{-0.1cm}
\label{fig:s3_s4_s5_fig}
\end{figure*}

\subsection{Homogeneous Accelerators}
We examined the Small homogeneous accelerator (S1) with system BW=16 GB/s across different tasks. As shown in \autoref{fig:exp_s1}, Herald-like and AI-MT-like mapper works rather well across four different tasks. The result form AT-MT-like shows that even though it is designed considering vision and language tasks, it also work adequately well in recommendation and even mix task. Likewise, Herald-like is designed for vision task and work well when applying to others. For optimization methods, they can reach similar performance as Herald-like and AI-MT-like. Note that, these optimization methods are not originally designed for this specific mapping task. However, they work adequately well working in the \framework framework. Overall, \alg outperforms others. \alg reach performance (geomean) 1.4x and 1.41x better than Herald and AI-MT, and (geomean) 1.6x better than other optimization methods.

\subsection{Heterogeneous Accelerators}
We examined the Small (S2) and Large (S4) heterogeneous accelerators across different tasks in \autoref{fig:exp_s2_4}. In the following results, we will focus on presenting the result of Mix task, since it is a more complex task and is a fair realistic use-case in nowadays' inference data centers~\cite{park2018deep, anderson2021first}. In \autoref{fig:exp_s2_4}, we also present Vision task result as baselines, since both Herald-like and AI-MT-like has Vision task as target. 

\textbf{Small Heterogeneous Accelerators (S2).} As shown in \autoref{fig:exp_s2_4}(a), Herald-like performs well, while AI-MT-like has comparatively lower performance. It shows the different characteristic of two algorithms. Herald-like is designed for heterogeneous accelerator while AI-MT-like is targeting homogeneous accelerator, which explains the performance difference. For optimization methods, many of them reach comparable performance to Herald-like, while PSO and CMA have lower performance (comparable to AI-MT-like). For a more complex Mix task (\autoref{fig:exp_s2_4}(b)), AI-MT-like has comparatively worse performance. Many optimization methods undergo lower performance, too. However, the two RLs methods stands out. Overall, \alg outperforms others in both tasks. \rev{B}y geomean, \alg is 2.3x better than Herald, 39.5x better than AI-MT, 13.4X better than optimization methods excluding RLs. RLs and \alg have compatible performance and \alg achieve slightly better result 1.01x better.

\textbf{Large Heterogeneous Accelerators (S4).}
In large accelerator, the mapping task becomes more complex since the design space of the mapping grows. As shown in \autoref{fig:exp_s2_4}(c)(d), Herald-like perform rather well in Vision task. However, at a more complex Mix task and a more complex large accelerator case, Herald-like starts to undergo lower performance. Many more basic optimization process cannot tackle the large and complex design space as well (\autoref{fig:exp_s2_4}(d)). However, RLs starts to shine and reach good performance. Overall, \alg outperforms others in both tasks. \rev{B}y geomean, \alg is 1.7x better than Herald, 52x better than AI-MT, 10x better than optimization methods excluding RLs, and 1.3x better than RLs. Note that the contribution of this work is both the framework \framework (which enables the other optimization methods) and algorithm \alg. Before this work, the best performing mapper in Large heterogeneous accelerator setting is Herald-like, which harvests only 20\% of maximum throughput enabled by \framework in Mix task, as shown in \autoref{fig:exp_s2_4}(d). 

\rev{\autoref{fig:scatter_plot} sketches how different methods explore and exploit, leading to their performance difference. For ablation study of sampling budget in different methods, we add a set of experiments, where we let all methods run until they converge, as shown in \autoref{fig:convergent_curve}.} 

\textbf{BW-limited Environment.} We examine the performance of Small accelerator at BW=16GB/s and Large accelerator at BW=256GB/s. However, in a heavy-loaded inference data center, the BW is a precious resource, where a big portion of it could also be occupied by other applications and leads to a more BW-limited environment. At a more BW-limited environment, mappers become crucial for smartly ordering the jobs to exploit the limited BW. We examine the effect of BW by a BW sweep in \autoref{fig:bw_sweep}. For both Small and Large accelerators, with the decrease of BWs, \alg stands out more obviously by reaching better relative performance. For example, in \autoref{fig:bw_sweep}(a), \alg is (geomean) 1.2x better than others when BW=16GB/s, but \alg is 1.6x better than other when BW=1GB/s. Similar observation can be made in \autoref{fig:bw_sweep}(b).

\subsubsection{Sub-accelerator Combinations}
In this experiment, we examine the performance change in different settings, S3 (Large Homogeneous), S4 (Large Heterogeneous), S5 (Large Heterogeneous, BigLittle) of the Large accelerator.

\textbf{Homogeneous versus Heterogeneous.}
In the following experiment, we discuss the performance implication of a homogeneous versus heterogeneous accelerator using \alg algorithm.
The LB-style sub-accelerators usually take larger runtime but lower BW requirements than HB-style in language and recommendation tasks, as shown in \autoref{fig:analysis_combine}(a). The jobs analysis in \autoref{fig:s3_s4_s5_fig}(a-b) reflect the fact that S4, in general, induces more no-stall latency but requires less BW than S3. Therefore, when BW is limited (BW=1), the heterogeneous setting enables accelerator to leverage the difference of BW requirement among sub-accelerators to relax the BW contention. Thus S4 reaches better performance than S3 at BW=1 in \autoref{fig:s3_s4_s5_fig}(c). However, when the BW is mostly sufficient (BW=256GB/S), the performance will reflect more of the behavior of the no-stall latency. Thus S3 reaches better performance.




\textbf{Bigs versus BigLittle.}
We consider an accelerator with a smaller setting, BigLittle (S5), comparing to Bigs (S3, S4). It is obvious when the BW budget is sufficient (BW=256GB/S), BigLittle will perform worse than both of the Bigs (S3, S4) as shown in \autoref{fig:s3_s4_s5_fig}(c), and can be verified by the jobs analysis in \autoref{fig:s3_s4_s5_fig}(a). However, BigLittle has smaller BW requirement because of its smaller sub-accelerator size, as shown in \autoref{fig:s3_s4_s5_fig}(b). Therefore, as shown in \autoref{fig:s3_s4_s5_fig}(c), when the BW is limited (BW=1), BigLittle (S5) with the least amount of resources reaches the best performance. This observation shows that in a multi-core heterogeneous system, in addition to making the compute cores more powerful (adding more compute resources to them), striking the balance between each cores is another key consideration for boosting the performance.

\begin{figure}[t]
\begin{center}
\includegraphics[width=1\linewidth]{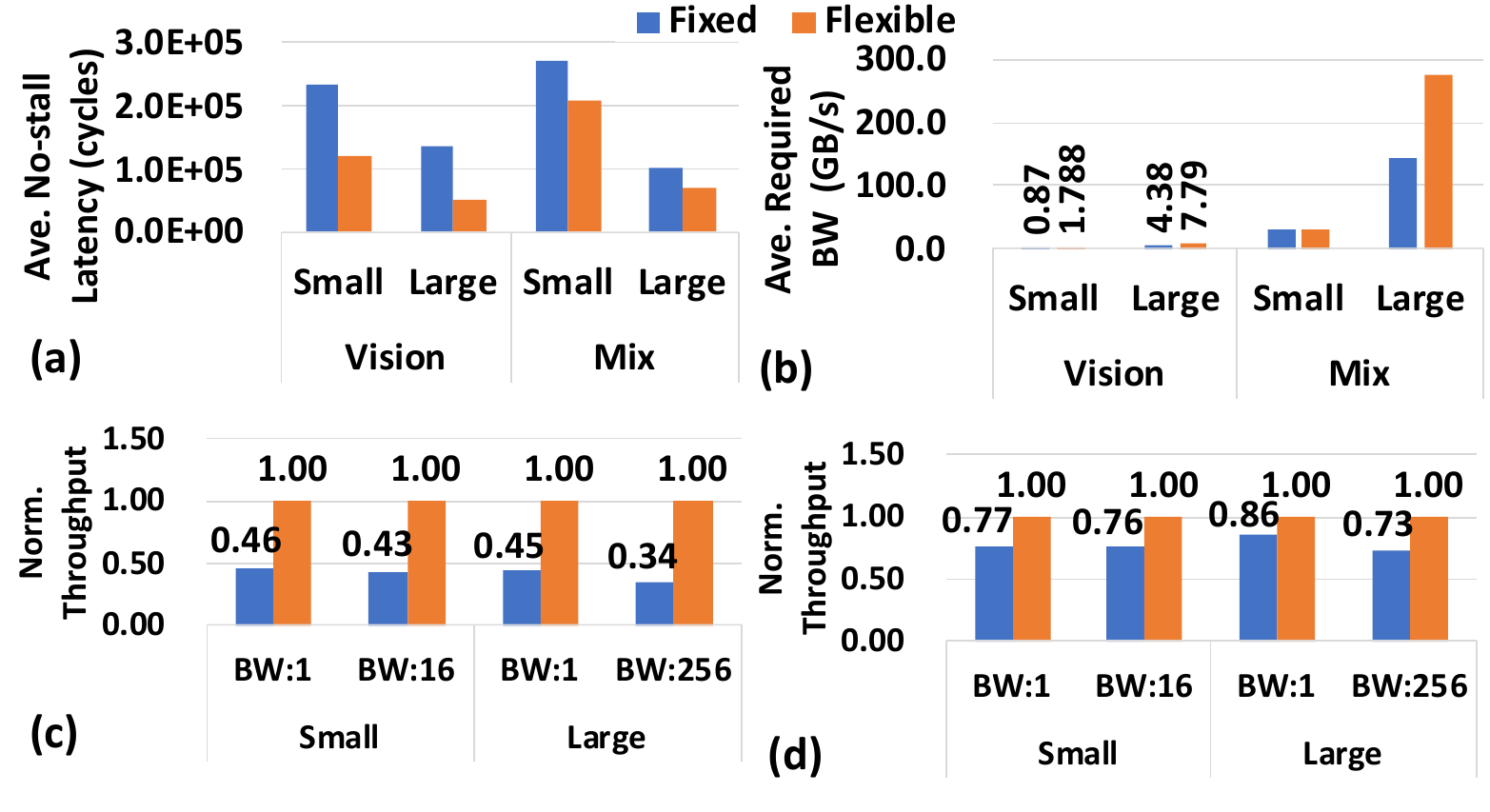}
\end{center}
\vspace{-0.3cm}
\caption{Jobs analysis of the averaged (a) per-job no-stall latency and (b) required BW of fixed and flexible PEs arrays. Performance evaluation of \alg with fixed or flexible PEs array on (c) Vision and (d) Mix. Throughput values are normalized by the value of flexible accelerator.}
\vspace{-0.2cm}
\label{fig:programmable_fig}
\end{figure}

\begin{figure}
\begin{center}
\includegraphics[width=1\linewidth]{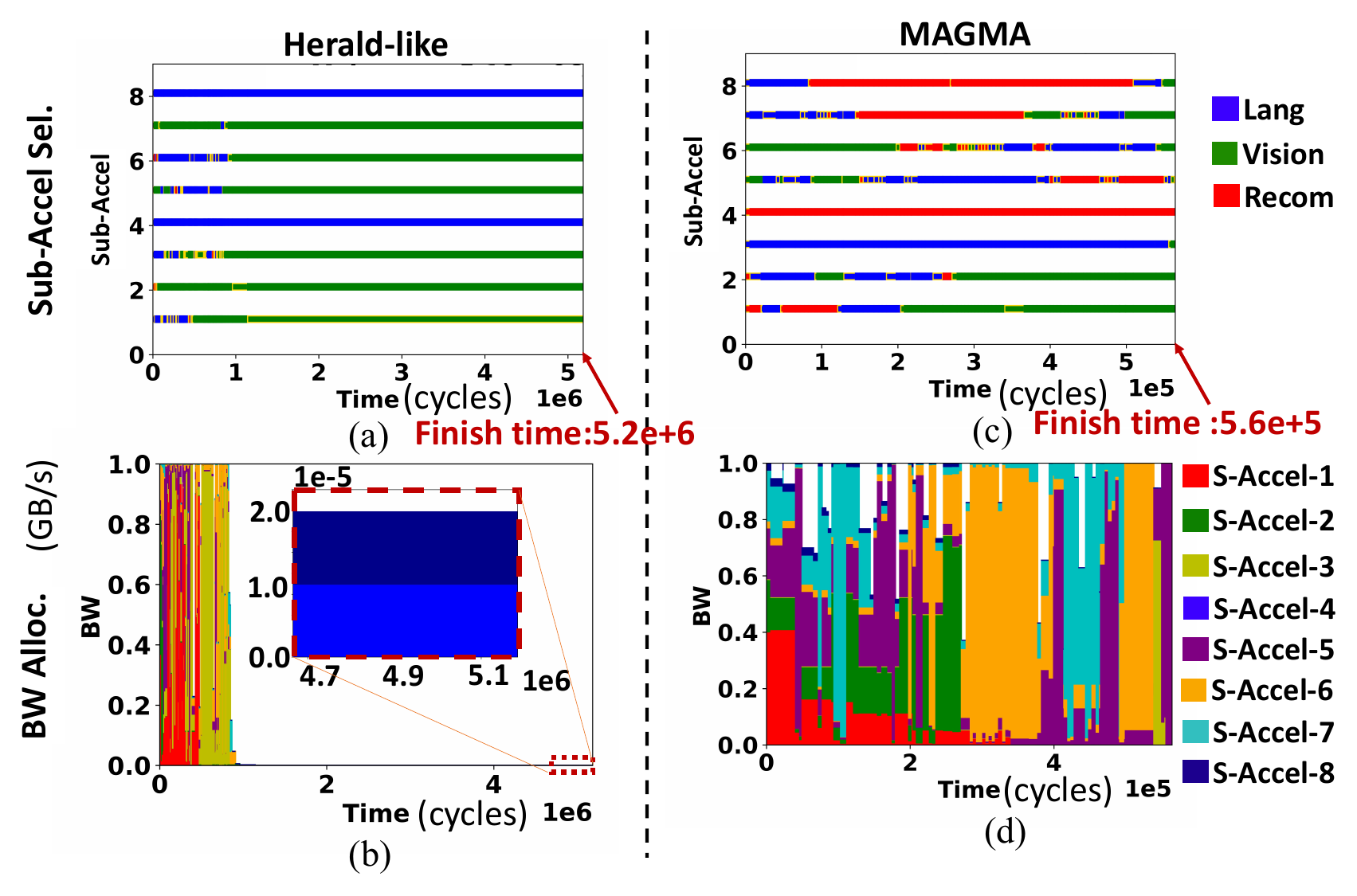}
\end{center}
\vspace{-0.3cm}
\caption{The visualization of found solution by Herald-like and \alg. (a)(c) shows the respective sub-accelerator allocations, and (b)(d) shows the respective BW allocations. (Mix task, S5, BW=1).}

\label{fig:bw_alloc_result}
\end{figure}

\begin{figure}[t]
\begin{center}
\includegraphics[width=1\linewidth]{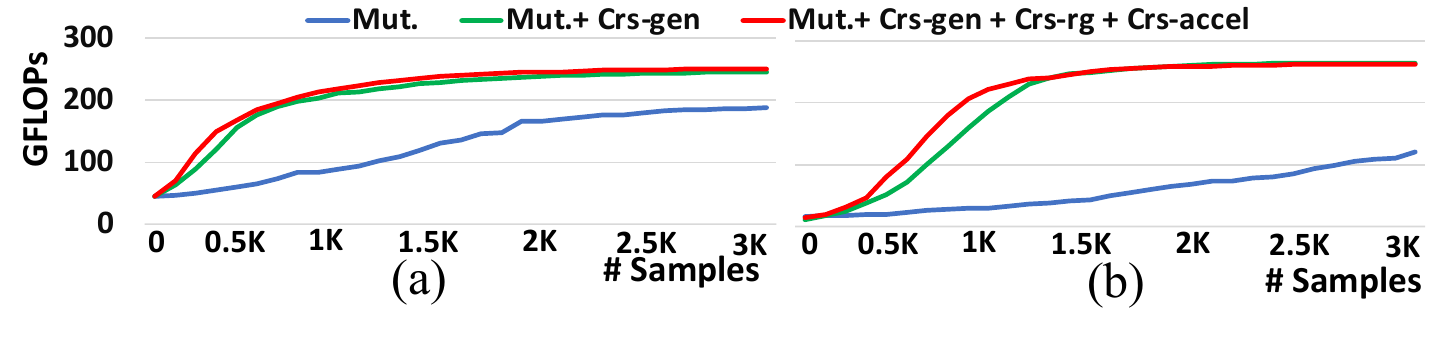}
\end{center}
\vspace{-0.4cm}
\caption{\rev{The convergence curve of \alg with three level of genetic operations: Mutation only, Mutation and Crossover-gen, and \alg with all four operators on (a) (Vision, S2, BW=16) and (b) (Mix, S3, BW=16). Mutation is the base operator in \alg. However, when \alg is restricted to only the mutation operator, the sample efficiency decreases seriously. We can see that adding Crossover-gen is essential for algorithm to find good solution within limited samples, and finally, Crossover-rg and Crossover-accel further help approach an optimum faster.}}

\label{fig:ablation}
\end{figure}

\begin{figure}
\begin{center}
\includegraphics[width=1\linewidth]{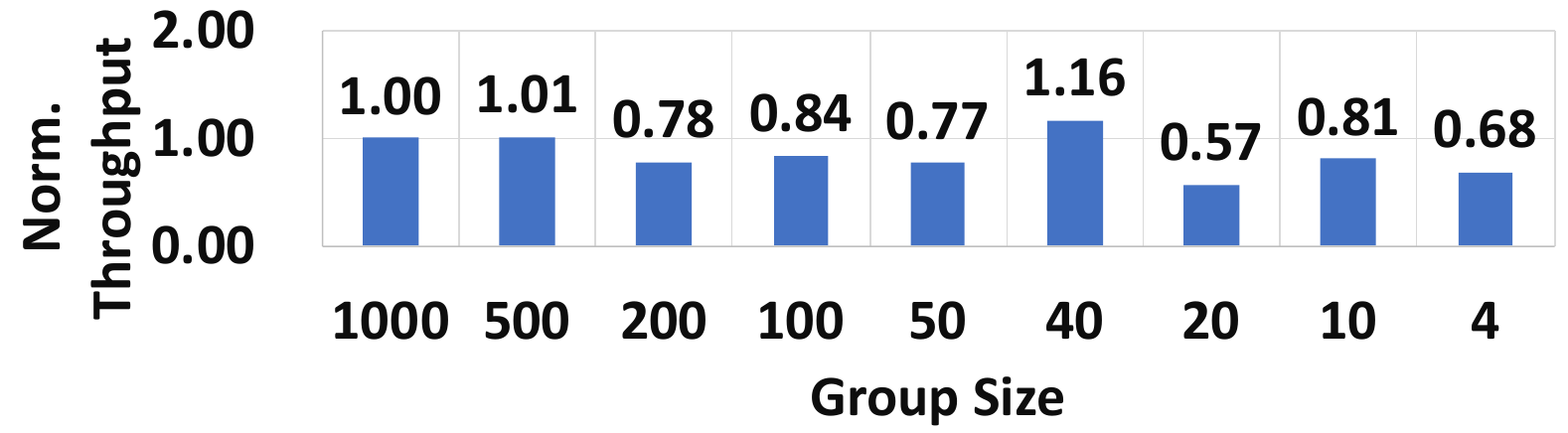}
\end{center}
\vspace{-0.3cm}
\caption{The reached performance of \alg given the same task and setting (Mix, S2, BW=16) with different group sizes. Throughput values are normalized by the value of \group size=1000,}

\label{fig:exp_groupsize}
\end{figure}

\begin{table}[t]

\centering
\caption{The performance of warm-start on (a) Mix, S4, BW=1. (b) The averaged performance across different tasks and different accelerator (S1-S6) under BW=1. All the values are normalized by the values of Trf-100-ep of each columns. Raw (highlighted in orange) is the throughput without warm-start. Trf-0-ep (highlighted in green) is warm-start and before further optimization. Trf-1-ep is warm-start with one epoch of optimization, and likewise for Trf-30-ep. Trf-100-ep (highlighted in blue) represents a full optimization process.}

\includegraphics[width=1\linewidth]{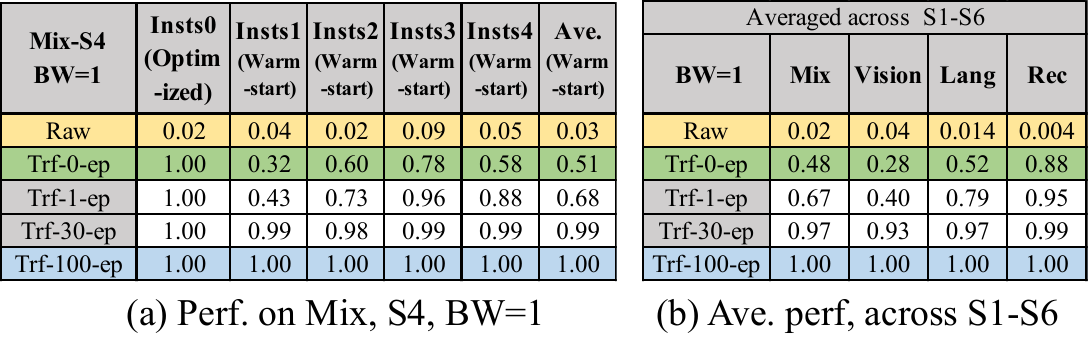}

\label{table:trf}
\end{table}


\subsection{Flexible Accelerator}
In this experiment, we consider accelerators where the PE array dimensions are configurable, such as  FPGAs~\cite{brainwave}, CGRA~\cite{maeri}, or programmable accelerators~\cite{bang201714,yin2018141,zheng2019ultra}, and demonstrate their performance by applying mapping found by \alg.

\textbf{Accelerator Configuration.}
We extend the setting of S1 (Small, fixed) and S3 (Large, fixed) to have flexible accelerators.
The number of PEs in the sub-accelerator are fixed (the same as in \autoref{table:setting}). However, the \textit{shape} of 2D PE arrays is flexible, that is we can configure the routing 
among the PEs. This enables the sub-accelerator to run
various dataflows or mappings~\cite{maeri}.
The maximum size of SLs are fixed as 1KB in each PE, and SGs are fixed as 2MB in each sub-accelerator. 

\textbf{Dataflow Strategy.} We pick the dataflow strategy 
of the sub-accelerator to maximize the utilization of the PEs array. In order to maximize the utilization, we will align the PEs array dimension to be the factor of the the parallelizing dimension of the tile as much as possible. For example if the parallelizing dimension of the tile is (2, 15), which is going to map over the y and x dimension of the PEs array with 16 PEs. The potential PE array shape could be 2$\times$8 while aligning to the factor of y dimension, or 3$\times$5, 5$\times$3, and 1$\times$15 while aligning to the factor of x dimension. We examine these combinations, evaluate their expected latency by the HW cost model, and pick the lowest latency one as our PE array configurations.


\textbf{Evaluations.}
From the performance analysis in \autoref{fig:programmable_fig}(a-b), we can observe that for both Vision and Mix tasks, \textit{flexible} outperforms \textit{fixed} in ave. per-job no-stall latency, owing to its ability to maximizing the utilization rate of the PEs array. However, it would also incur higher BW requirement. It is because the flexible mapping we found is to maximize the PE utilization rate, which also increases the number of data to fetch per tile to keep PEs busy.

From all scenario in \autoref{fig:programmable_fig}(c-d), \textit{flexible} outperforms \textit{fixed}.
The results conclude that with flexible accelerators (ASIC of FPGA), we could further increase the accelerator performance without providing additional compute HW resources (PEs) if the accelerators (or sub-accelerators) have configurable PEs array shape. 

\subsection{More about \alg Algorithm}
\label{sec:eval_deepdive}

\textbf{Analysis of found solutions.}
To understand the effect of different mapping, we show the detailed sub-accelerator selection and the corresponding BW allocation results of mapping found by two of the mappers, Herald-like and \alg. We showcase what is actually happening on the accelerator in an execution duration of one \group of jobs in \autoref{fig:bw_alloc_result}. We found that \alg can distribute the BW-intensive jobs (Recommendation, Language) across the runtime to balance the BW requirement (\autoref{fig:bw_alloc_result}(c-d)). In contrast, Herald-like (\autoref{fig:bw_alloc_result}(a-b)) tries to use BW intensively at the beginning, which causes BW competition. Finally, it causes longer finish time of a \group of jobs comparing to \alg. 

\textbf{Warm-start of \alg.}
\label{sec:transfer_exp}
In the following, we show the usefulness of warm-start technique in \alg. 
In the experiments, we use \alg to optimize on a \group of jobs, Insts0. Then, we test, and optimize on the other four different \group of jobs. \autoref{table:trf}(a) shows that by directly applying previous knowledge (Trf-0-ep), we could achieve \textbf{16x better performance} than the usual starting points, randomly initialization (Raw). By warm-start followed by one epoch/ step of optimization (Trf-1-ep), we could already receive 93\% of the expected performance gain of a full optimization (Trf-100-ep).
We execute the same experiment for different types of tasks and for different setting (S1-S6) (\autoref{table:trf}(b)). 
We can observe for BW-intensive tasks, Language and Recommendation, the previous knowledge become more important, and therefore the performance gain from the warm-start become significant. Overall, by warm-start and before further optimization is run (Trf-0-ep), \alg can achieve \textbf{7.4x} to \textbf{152x} better performance than the the usual starting points (Raws).


\rev{\textbf{Ablation Study of Operators.} \autoref{fig:ablation} shows the importance of the four designed operators. We can see that we could achieve the best sampling efficiency when all four operators are included.} 

\textbf{Ablation Study of Group Size.} Throughout the evaluation, we use the benchmark with a set \group size of 100. It establishes a fair comparisons of the performance of different mappers. Also, in practice, \group size is often a pre-defined system parameter as the formulated benchmark.
However, 
a larger (or smaller) \group size is also valid. We execute a \group size sweep in \autoref{fig:exp_groupsize} using \alg algorithm. It tells that increasing or decreasing the \group size does not affect the overall performance drastically. However, a too small \group size (e.g., 4) will lead to lower performance.

\section{Related Works}
\label{sec:related}

\textbf{Mapping DNN Jobs on Single Accelerator.}
Several mappers have been proposed for the problem of mapping a single DNN layer efficiently on an accelerator. These include manual-designed mapping search~\cite{shen2017maximizing, lu2017flexflow}, heuristic-based mapping search~\cite{timeloop, stoutchinin2019optimally, systolic_mapping, cong_fpga} and optimization/ML methods~\cite{confx, gamma, mindmapping, cosa, suda2016throughput}. These works fall 
within the local mapping phase within individual accelerator cores.

\textbf{Multi-tenant Mapping for DNN Accelerators.}
Prophet~\cite{prophet} builds a runtime prediction model for multi-tenancy on GPU. AI-MT~\cite{aimt} develops a heuristic for DNN job mapping for multi-PE arrays. Prema~\cite{prema} explores preemptive multi-tenancy on a NPU. Herald~\cite{herald} and Planaria~\cite{planaria} use manual-designed mapping for assigning jobs to sub-accelerators or reconfigurable PEs array. SCARL~\cite{scarl} utilizes RL for the mapping problem. In this work, we target multi-DNNs mapping and compared \alg against prior arts~\cite{aimt, herald}, black-box optimizations~\cite{tbpsa, ga, de, cma, pso_paper}, and RLs~\cite{a2c,ppo2}.

\textbf{Multi-tenant Scheduling for CPUs and GPUs.}
Multi-tenancy has been investigated for decades for multi-tasking on a single CPU and job ordering in CPU clusters~\cite{zaharia2009job,sherwani2004libra} or in GPUs ~\cite{beisel2011cooperative, joo2014resource}. 
GAs~\cite{correa1999scheduling,hou1994genetic,singh1996mapping,shroff1996genetic,wang1996genetic}, PSO~\cite{wu2010revised}, CMA-ES~\cite{emadi2017task}, and other optimizations have also been used.
Some works leverage RL for jobs ordering over clusters such as DeepRM ~\cite{deeprm}, Decima~\cite{decima} and Thamsen \textit{et al.}~\cite{thamsen2017scheduling}. However, they presume a unified abstraction of the underlying cluster, where heterogeneity of the system is not considered.

\section{Conclusions}
This work presents a mapping optimizer for multi-tenant DNN accelerators.
The key takeaways are as follows. (i) Heuristic and optimization methods have been used successfully for the design space of local-mapping (i.e., dataflow design). However, global-mapping forms a new drastically different search space. A new mapper for global mapping is needed for upcoming platforms (\autoref{table:highlevel_comparison}). (ii) The search space for this (global-) mapping is extremely enormous.
The search sample-efficiency of baseline optimization methods is not sufficient to find optimized solutions.
(iii) We develop an optimization algorithm called \alg that
customizes its exploration momentum and mechanism (genetic operators in this work) for the target search space and outperform the existing related works and other well-known optimization methods.
\section*{Acknowledgment}
This work was supported by NSF Award 1909900.
We thank Suvinay Subramanian for valuable technical discussions and helping us with positioning this work.


\bibliographystyle{IEEEtranS}
\bibliography{main}

\end{document}